\documentclass[11pt,a4paper]{article}
\pdfoutput=1

\usepackage{jheppub}

\usepackage{amsmath}
\usepackage{slashed}
\usepackage{caption}
\usepackage[utf8]{inputenc}
\usepackage{amsmath, amsfonts, amssymb}
\usepackage{shuffle}
\usepackage{mathtools}
\usepackage{physics}
\usepackage{subcaption}
\usepackage{booktabs,array}

\usepackage{cleveref}

\usepackage{xcolor}
\usepackage{listings}
\usepackage{algorithm}
\usepackage{upquote} 

\def\eps{\epsilon}

\def\trfive{\tr_5}

\newcommand{\ii}{\,\mathrm{i}\,}

\newcommand{\p}[1]{(\ref{#1})}

\newcolumntype{C}[1]{>{\centering\arraybackslash}m{#1}}

\makeatletter
\newcommand\footnoteref[1]{\protected@xdef\@thefnmark{\ref{#1}}\@footnotemark}
\makeatother

\definecolor{codegray}{rgb}{0.5,0.5,0.5}
\definecolor{codepurple}{rgb}{0.58,0,0.82}
\definecolor{backcolour}{rgb}{0.95,0.95,0.92}

\title{Pentagon Functions for One-Mass Planar Scattering Amplitudes}

\author[a]{Dmitry Chicherin,}
\emailAdd{chicherin@lapth.cnrs.fr} 

\author[b]{Vasily Sotnikov,}
\emailAdd{sotnikov@mpp.mpg.de}

\author[c]{Simone Zoia}
\emailAdd{simone.zoia@unito.it}

\affiliation[a]{
LAPTh, Universit\'e Savoie Mont Blanc, CNRS, B.P. 110, F-74941 Annecy-le-Vieux, France
}

\affiliation[b]{
Max Planck Institute for Physics (Werner Heisenberg Institute), D–80805 Munich, Germany
}

\affiliation[c]{
Dipartimento di Fisica and Arnold-Regge Center, Universit\`{a} di Torino,
and INFN, Sezione di Torino, Via P. Giuria 1, I-10125 Torino, Italy
}

\abstract{
  We present analytic results for all planar two-loop Feynman integrals contributing to five-particle scattering amplitudes with one external massive leg.
  We express the integrals in terms of a basis of algebraically-independent transcendental functions, which we call one-mass pentagon functions.
  We construct them by using the properties of iterated integrals with logarithmic kernels. The pentagon functions are manifestly free of unphysical branch cuts,
  do not require analytic continuation,
  and can be readily evaluated over the whole physical phase space of the massive-particle production channel.
  We develop an efficient algorithm for their numerical evaluation and present a public implementation suitable for direct phenomenological applications.
}

\preprint{LAPTH-041/21, MPP-2021-182}

\begin{document}

\setcounter{tocdepth}{2}

\maketitle
\flushbottom

%\listoftodos

\section{Introduction}
\label{sec:Introduction}

Electroweak boson production in association with photons and jets plays a key role in the physics program of hadron colliders such as the Large Hadron Collider (LHC).
It provides a window into understanding the electroweak symmetry breaking mechanism and constitutes a prominent irreducible background in new-physics searches.
Our ability to correctly interpret the experimental data relies to a large extent on the availability of precise theoretical predictions for this class of processes.
Generally the knowledge of the next-to-next-to-leading order (NNLO) in the perturbative expansion in the strong coupling constant is highly desirable in order to harness the wealth of data to be collected.
To access the important $W/Z b\bar{b}$ and $W/Z\gamma\gamma$ production processes, as well as $W/Z$ + multijet associated production, the calculation of $2\to3$ NNLO corrections is required. The latter represents a major challenge and remains on the forefront of modern computational methods' capabilities.

The remarkable progress in the computation of massless two-loop five-particle scattering amplitudes in the last few years
\cite{Gehrmann:2015bfy,Badger:2018enw,Abreu:2018aqd,Chicherin:2018yne,Chicherin:2019xeg,Abreu:2019rpt,%
Abreu:2018zmy,Abreu:2018jgq,Abreu:2019odu,Badger:2019djh,Abreu:2020cwb,Chawdhry:2020for,Hartanto:2019uvl,Caron-Huot:2020vlo,Agarwal:2021grm,Abreu:2021oya,%
Chawdhry:2021mkw,Agarwal:2021vdh,Badger:2021imn,DeLaurentis:2020qle} 
has recently culminated in the first NNLO theoretical predictions for $2\to3$ processes \cite{Chawdhry:2019bji,Kallweit:2020gcp,Chawdhry:2021hkp,Czakon:2021mjy,Badger:2021ohm}. 
This success was enabled by the availability of analytic results for the required two-loop Feynman integrals~\cite{Papadopoulos:2015jft,Gehrmann:2018yef,Abreu:2018rcw,Chicherin:2018mue,Abreu:2018aqd,Chicherin:2018old,Chicherin:2020oor,Gehrmann:2015bfy,Badger:2019djh}.
In particular, it was crucial to express the latter in terms of judicious bases of transcendental functions \cite{Gehrmann:2018yef,Chicherin:2020oor}.
On the one hand, these bases make the analytic structure of scattering amplitudes manifest, enable the analytic cancellation of ultraviolet and infrared divergences, and facilitate the derivation of compact representations.
On the other hand, they are amenable for efficient and stable numerical evaluation suitable for phenomenological applications \cite{Chicherin:2020oor}.
In this paper, we construct a similar basis of transcendental functions for the planar two-loop five-particle scattering amplitudes with a single external massive particle.

The planar one-mass two-loop five-point Feynman integrals are currently the subject of extensive study. 
The four-point integrals with two external masses were computed in refs.~\cite{Henn:2014lfa,Papadopoulos:2014hla,Gehrmann:2015ora}. As for the genuine five-point Feynman integrals, bases of independent integrals ---~alias master integrals~--- satisfying systems of canonical differential equations (DEs) \cite{Henn:2013pwa} have been constructed in ref.~\cite{Abreu:2020jxa}. Their numerical evaluation was achieved by solving the DEs in terms of generalized power series expansions~\cite{Moriello:2019yhu,Hidding:2020ytt,Abreu:2020jxa}.
Expressions for the master integrals in terms of multiple polylogarithms (MPLs)~\cite{Goncharov:1998kja,Remiddi:1999ew,Goncharov:2001iea} have been derived in refs.~\cite{Papadopoulos:2015jft,Canko:2020ylt,Syrrakos:2020kba}.
In ref.~\cite{Badger:2021nhg}, a basis for the subset of transcendental functions contributing to completely color ordered amplitudes has been constructed.
Similarly to ref.~\cite{Abreu:2020jxa}, their numerical evaluation was performed through series expansion of their DEs.
These developments have led to the first analytic calculations of two-loop five-particle amplitudes with an external massive leg, for the production of a $W$ and a Higgs bosons associated with a pair of bottom quarks \cite{Badger:2021nhg,Badger:2021ega},
as well as of the two-loop four-point form factor of a length-3 half-BPS operator in planar $\mathcal{N}=4$ sYM \cite{Guo:2021bym}.
First results for the non-planar integrals have also been reported recently \cite{Papadopoulos:2019iam,Abreu:2021smk,Liu:2021wks}.

In this paper we construct a basis of transcendental functions that is sufficient to express any planar two-loop corrections for scattering cross sections involving four massless and one massive particle.
We call the functions in this basis \textit{one-mass pentagon functions}.
We follow the approach of ref.~\cite{Chicherin:2020oor}. We start from the canonical DEs of ref.~\cite{Abreu:2020jxa} and generalize them to all permutations of the external massless legs.
We compute the initial values for the DEs using the expressions of the master integrals in terms of MPLs from refs.~\cite{Canko:2020ylt,Syrrakos:2020kba}.
We write the solutions order by order in the dimensional regularization parameter $\epsilon$ in terms of iterated integrals \cite{Chen:1977oja} with logarithmic kernels.
The solutions are pure functions of uniform transcendental weight \cite{Arkani-Hamed:2010pyv,Henn:2013pwa} and satisfy a shuffle algebra, which we use to linearize and eliminate all algebraic relations in the space of functions generated by the DEs.
The algebraic independence and closure over momenta permutations of the one-mass pentagon functions is very advantageous for the modern workflow to compute scattering amplitudes analytically,
which is based on the application of finite field arithmetic and functional reconstruction \cite{vonManteuffel:2014ixa,Peraro:2016wsq}.
Indeed, writing all permutations of the Feynman integrals in terms of the same basis of transcendental functions gives access to simplifications of the rational coefficients that would otherwise be missed.
We find explicit representations of the basis functions up to weight two in terms of logarithms and dilogarithms, and up to weight four in terms of one-dimensional integrals.
All the expressions are branch-cut free within the physical scattering region of the massive-particle production channels.
We thus demonstrate that the constructive algorithm of ref.~\cite{Chicherin:2020oor} can be straightforwardly generalized beyond the purely massless kinematics.
The numerical evaluation of the weight three and four pentagon functions on the other hand is significantly more challenging compared to their massless counterpart.
This is due to the non-trivial geometry of the phase space, which we study in detail, and to the presence of non-linear spurious singularities in the one-dimensional integral representations.
Nevertheless, we design an algorithm for the numerical evaluation that is fast and stable across the whole phase space of the $2\to3$ massive-particle production channels.
We achieve performance comparable to that of the massless pentagon functions from ref.~\cite{Chicherin:2020oor},
and provide a public implementation which meets the demanding requirements of phenomenological applications. 

Compared to a more traditional approach of attempting to express the master integrals in terms of MPLs, our method is rather insensitive to the presence of square roots in the symbol alphabet.
Even in cases where the canonical ``$\epsilon \, \dd{\log}$"-form of DEs can be obtained,
it can be difficult, if not impossible, to find MPL solutions if the symbol alphabet contains non-rationalizable square roots \cite{Heller:2019gkq,Brown:2020rda,Canko:2020ylt,Kreer:2021sdt,Bonetti:2020hqh}.
Very importantly, even when these attempts did succeed, it was frequently observed that the numerical evaluation in the physical scattering regions is very challenging \cite{Papadopoulos:2015jft,Gehrmann:2018yef,Canko:2020ylt,Duhr:2021fhk}.
In addition, contrary to na\"ive expectations, the physical properties may even be more obscured due to the presence of spurious branch cuts, which are instead absent in the iterated integral representation.
Our approach elegantly bypasses these issues, and therefore can be a useful alternative for cases when the integration of the DEs in terms of MPLs is impractical or impossible.

The rest of the paper is structured as follows. We begin in section~\ref{sec:Kinematics} by presenting our notation, discussing the kinematics, and defining the $2\to3$ physical scattering regions. In section~\ref{sec:DE} we define the planar families of five-particle integrals with one external massive leg up to two loops, and discuss how to express the master integrals in terms of Chen's iterated integrals through the canonical DEs they satisfy. In section~\ref{sec:FunctionBasis} we discuss the construction of our function basis. First, in section~\ref{sec:BasisConstruction} we use the properties of Chen's iterated integrals to construct systematically a basis of algebraically independent functions to express all permutations of the master integrals up to the order in $\eps$ required for two-loop corrections to cross sections. Then, in section~\ref{sec:explicit-repr} we design explicit expressions for the basis functions which are well suited for their numerical evaluation, and show that they are well defined in the chosen physical scattering region. We discuss the implementation in a \texttt{C++} library of routines for the numerical evaluation of our function basis in section~\ref{sec:Implementation}.
We draw our conclusions in section~\ref{sec:Conclusions}. 
We also include a number of appendices,
where we discuss in detail the physical phase-space geometry and the positivity properties of the symbol alphabet.

\section{Scattering kinematics}
\label{sec:Kinematics}

In this section we discuss the scattering kinematics and introduce certain quantities which will play an important role in the rest of the paper. We consider the scattering of five particles, four massless and one massive.
We take their momenta $p_i$ to be outgoing, and the momentum conservation reads
\begin{align}
\sum_{i=1}^5 p_i = 0 \,.
\end{align}
Following the notation of refs.~\cite{Abreu:2020jxa,Abreu:2021smk}, we assume that momentum $p_1$ is massive, while the others are massless, 
\begin{align}
p_1^2 \neq 0 \,, \qquad \qquad p_i^2=0 \quad \forall i \in \{2,3,4,5\} \,.
\end{align}
We parametrize the scattering kinematics by six Mandelstam invariants,
\begin{align}\label{eq:Mand}
X = \left( p_1^2, s_{12}, s_{23}, s_{34}, s_{45}, s_{51} \right)\,, \qquad s_{ij} = (p_i + p_j)^2 \,,
\end{align}
and the sign of the imaginary part of the parity-odd invariant
\begin{align} \label{eq:tr5}
\trfive := 4 \ii \, \varepsilon_{\mu \nu \rho \sigma}p_1^\mu p_2^\nu p_3^\rho p_4^\sigma\,,
\end{align}
where $\varepsilon_{\mu \nu \rho \sigma}$ is the fully antisymmetric Levi-Civita symbol.
The square of $\trfive$ is algebraically related to the Mandelstam invariants in \cref{eq:Mand} through the Gram determinant of four external momenta\footnote{See the definition in \cref{app:GramDet}.} as
\begin{align}
  \label{eq:tr5Delta5}
  \trfive^2 = \Delta_5 \coloneqq  16 \, G(p_1 ,p_2,p_3, p_4)   \,, 
\end{align}
with
\begin{equation} \label{eq:Delta5}
\begin{aligned}
\Delta_5 = \bigl(s_{12} (s_{15} - s_{23}) - s_{15} s_{45} + s_{34} & (s_{23} + s_{45}-p_1^2)\bigr)^2 \\  
 & + 4 s_{23}  s_{34} s_{45}  (s_{12} + s_{15} - s_{34}-p_1^2 ) \,.
\end{aligned}
\end{equation}
We emphasize that, despite the relation in \cref{eq:tr5Delta5}, the sign of $\Im \trfive$ is required to fully specify a point in the scattering phase space.
The sign of $\Im \trfive$ is changed under parity conjugation or permutations of external momenta, whereas $\sqrt{\Delta_5}$ stays unchanged.
Therefore one should be careful to distinguish these objects.

The other three Gram determinants that give rise to square roots which are relevant for the analytic structure of the five-particle one-mass scattering amplitudes are 
\begin{align}
\Delta_3^{(1)} := - 4 \, G\left(p_1, p_2+p_3 \right) = \lambda\left(p_1^2, s_{23}, s_{45} \right) \,, \\
\Delta_3^{(2)} := - 4 \, G\left(p_1, p_2+p_4 \right) = \lambda\left(p_1^2, s_{24}, s_{35} \right) \,, \\
\Delta_3^{(3)} := - 4 \, G\left(p_1, p_3+p_4 \right) = \lambda\left(p_1^2, s_{25}, s_{34} \right) \,, \label{Deltalambda}
\end{align}
where $\lambda$ denotes the K\"allen function,
\begin{align}
\lambda(a,b,c) := a^2 + b^2 + c^2 - 2 a b - 2 a c - 2 b c \,. \label{eq:Kaellen}
\end{align}
The set $\{\Delta_3^{(1)},\Delta_3^{(2)},\Delta_3^{(3)}\}$ is closed under the permutations of external massless momenta $S_4$, namely the $\Delta_3^{(i)}$'s transform into each other under 
the action of $\sigma\in S_4$.
In this paper we denote by 
\begin{align}
\sigma = \left(\sigma_2\sigma_3\sigma_4\sigma_5\right), \qquad \sigma_i\in\{2,3,4,5\} \,,
\end{align}
the permutation $\sigma \in S_4$ such that $\sigma(i) = \sigma_i$.
The action of $\sigma$ on the external massless momenta is given by
\begin{align}
\sigma : \left(p_1,p_2,p_3,p_4,p_5\right) \rightarrow \left(p_1 ,p_{\sigma(2)},p_{\sigma(3)},p_{\sigma(4)},p_{\sigma(5)}\right) \,,\label{sigma}
\end{align} 
and the permutation $\sigma$ of a generic function $f$ of the external kinematics ---~e.g.\ a Feynman integral or a scattering amplitude~--- is given by
\begin{align}
\left(\sigma \circ f \right)\left(s_{ij}, \trfive\right) = f\left(s_{\sigma(i) \sigma(j)}, \sigma \circ \trfive\right)  \,. \label{sigmaf}
\end{align}
From the definition~\eqref{eq:tr5} it follows that
\begin{align}
\sigma \circ \trfive = \mathrm{sign}(\sigma) \, \trfive \,,
\end{align}
where $\mathrm{sign}(\sigma)$ is the signature of the permutation $\sigma$. 

The Feynman integrals are multi-valued functions of the kinematics with a very intricate analytic structure. It is therefore fundamental to specify the domain of the kinematic variables $X$. 
We consider the physical scattering regions associated with the production of one massive and two massless particles,
\begin{align} \label{eq:process}
i\, j \rightarrow 1\, k\, l \,, 
\end{align}
where $i,j,k,l$ take distinct values in $\{2,3,4,5\}$. These regions are relevant for instance for the production of a massive vector boson in association with two jets at a hadron collider. 
Without loss of generality we assume that the incoming momenta are $p_4$ and $p_5$, i.e.\ we focus on the $s_{45}$ channel.
The $s_{45}$ channel is defined by requiring that the momenta $\{p_i\}_{i=1}^5$ correspond to physical configurations of the scattering process $45\to 123$, i.e.\ they are real and are associated with real scattering angles and positive energies. These constraints translate into the following inequalities for the scalar products of the momenta,
\begin{equation} \label{eq:pipjConstraints}
\begin{aligned}
& p_4 \cdot p_5 > 0 \,,~p_1 \cdot p_2 > 0 \,,~ p_1 \cdot p_3 > 0 \,,~p_2 \cdot p_3 > 0 \,,~\\
 p_4 \cdot p_1 < 0 \,,~\ & p_4 \cdot p_2 < 0 \,,~p_4 \cdot p_3 < 0 \,,~p_5 \cdot p_1 < 0 \,,~p_5 \cdot p_2 < 0 \,,~p_5 \cdot p_2 < 0 \,,
\end{aligned}
\end{equation}
as well as the following Gram-determinant inequalities~\cite{Byers:1964ryc,POON1970509,Byckling:1971vca},
\begin{subequations} \label{eq:GramDetConstraints}
  \begin{gather} 
    G(p_1) > 0\,, \qquad G(p_i) = 0,  \quad \forall i \neq 1 \,, \\
    G(p_i,p_j) < 0 \,, \qquad G(p_i,p_j,p_k) > 0 \,, \qquad G(p_i,p_j,p_k,p_l) < 0 \,,
  \end{gather}
\end{subequations}
where $i,j,k,l$ take distinct values in $\{1,2,3,4,5\}$.
We give a thorough discussion of the role of the Gram determinants in appendix~\ref{app:GramDet}.
Here we content ourselves with noting that the only non-vanishing Gram determinant involving only one momentum is $G(p_1)=p_1^2$, which we assume to be positive, and that all the Gram determinants involving four momenta are proportional to $\Delta_5$. The negativity of $\Delta_5$ can be understood as a consequence of the reality of the momenta, which implies that $\trfive$ is purely imaginary (see \cref{eq:tr5}) and hence, through \cref{eq:tr5Delta5}, that $\Delta_5 < 0 $. Putting together all the inequalities above and simplifying them (see appendix~\ref{app:GramDet}) gives the following definition of the $s_{45}$ scattering channel $45 \to 123$, which we label by $\mathcal{P}_{45}$:%
\footnote{
  \label{footnote:degenerate-momenta}
  Strictly speaking, the inequalities in \cref{eq:GramDetConstraints} exclude certain subvarieties of the physical phase space where the external momenta conspire to span a vector space of dimension lower than four.
  The points in these subvarieties correspond to loci of some Gram determinants vanishing, hence a subset of the boundary of $\mathcal{P}_{45}$ corresponds to valid (degenerate) scattering configurations.
  Since these subvarieties have measure zero in the phase space, for the purpose of this paper we can ignore them. 
}
\begin{equation} \label{eq:P45}
\begin{aligned}
\mathcal{P}_{45} = \bigl\{ X \ \bigl| \ &  p_1^2 > 0 \,,~p_2^2 = 0\,,~p_3^2 = 0\,,~p_4^2 = 0\,,~p_5^2 = 0\,,~\Delta_5 < 0 \,, \\
  & s_{12} > p_1^2\,,~s_{13} > p_1^2 \,,~s_{23}>0 \,,~s_{45}>p_1^2 \,,  \\ 
&  s_{24} < 0\,,~s_{34} < 0 \,,~s_{25} < 0 \,,~s_{35} < 0 \,,~s_{14} < 0 \,,~s_{15} < 0 \bigr\} \, .
\end{aligned}
\end{equation}
In other words, given a set of Mandelstam invariants~\p{eq:Mand} satisfying the inequalities~\p{eq:P45}, there is a corresponding configuration of particle momenta describing the scattering process $45 \to 123$.\footnote{It is understood that the particles have positive energies in a physical scattering process, and that the momentum configuration is unique up to an orthochronous Lorentz transformation.} 
Moreover, since $\trfive$ is purely imaginary for real momenta, the scattering region $\mathcal{P}_{45}$ is separated into two halves, corresponding to $\Im \, \trfive > 0$ and $\Im \, \trfive < 0$,
\begin{align} \label{eq:P45pm}
\mathcal{P}_{45}^{+} = \left\{ X \in \mathcal{P}_{45} \ \bigl| \ \Im \, \trfive > 0 \right\} \,, \qquad
\mathcal{P}_{45}^{-} = \left\{ X \in \mathcal{P}_{45} \ \bigl| \ \Im \, \trfive < 0 \right\} \,.
\end{align}
Each of the halves is a connected region in the momentum space~\cite{Byers:1964ryc}. The two halves are mapped onto each other by a parity transformation.

Apart from the massive-particle production channels~\eqref{eq:process}, other possible physical channels are the decay into four particles ($1 \to 2345$) and three-particle production with a single massive particle in the initial state ($1 j \to klm$ where $j$, $k$, $l$, $m$ take distinct values in $\{2,3,4,5\}$).
Here we focus on the case of massive-particle production, which is the most important for the hadron-collider phenomenology.

\section{Integral families and differential equations}
\label{sec:DE}

The modern approach to computing scattering amplitudes analytically relies on the fact that they can be expressed in terms of linear combinations of scalar Feynman integrals. The latter are then grouped into families based on their propagator structure.
The advantages are that within each family only a finite number of integrals are linearly independent ---~typically much fewer than the Feynman diagrams contributing to the amplitudes~--- and that it is possible to express the amplitudes in terms of them algorithmically.
These linearly independent integrals, called master integrals in the literature, constitute the basis of the linear span of all the integrals in the family. Crucially, the master integrals satisfy a system of first-order differential equations (DEs)~\cite{Kotikov:1990kg,Bern:1993kr,Remiddi:1997ny,Gehrmann:1999as,Henn:2013pwa}, which constitutes a powerful tool to compute them analytically. In this section we discuss the planar Feynman integral families relevant for the scattering of one massive and four massless particles at one and two loops with internal massless propagators. We closely follow ref.~\cite{Abreu:2020jxa} and lift their results to all $S_4$ permutations of the external massless particles. We begin by defining the integral families. Then we discuss the DEs satisfied by the master integral bases proposed in ref.~\cite{Abreu:2020jxa}, and show how they can be solved in terms of Chen's iterated integrals~\cite{Chen:1977oja}. We finish the section by discussing how we computed the initial values necessary to fix uniquely the solution of the DEs. The resulting expression of the master integrals in terms of iterated integrals is then the starting point of the construction of the function basis, which we discuss in section~\ref{sec:FunctionBasis}.

\subsection{Integral families}
\label{sec:IntegralFamilies}

\begin{figure}[t!]
\centering
\begin{subfigure}{.5\textwidth}
    \centering
    \includegraphics[width=.7\textwidth]{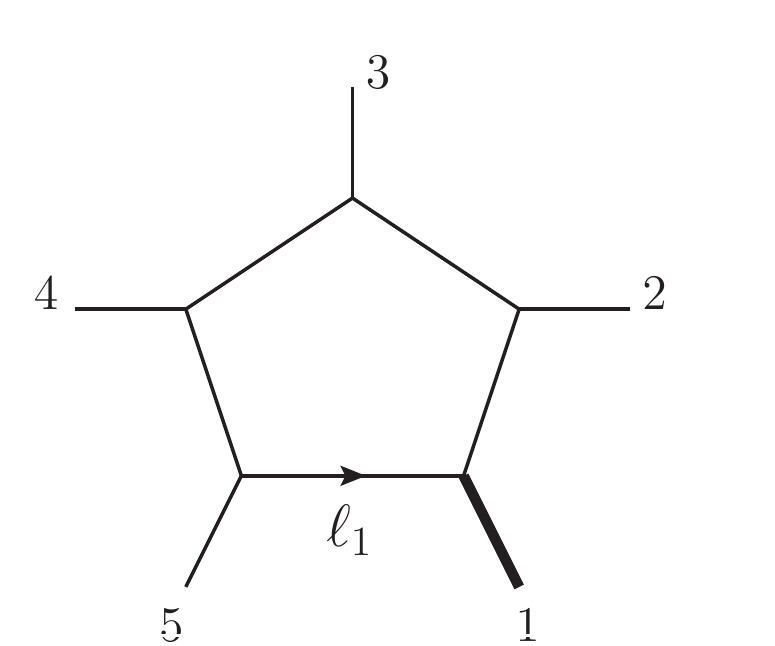}
    \caption{${\rm 1L}$}
    \label{fig:1L}
\end{subfigure}%
\begin{subfigure}{.5\textwidth}
    \centering
    \includegraphics[width=.8\textwidth]{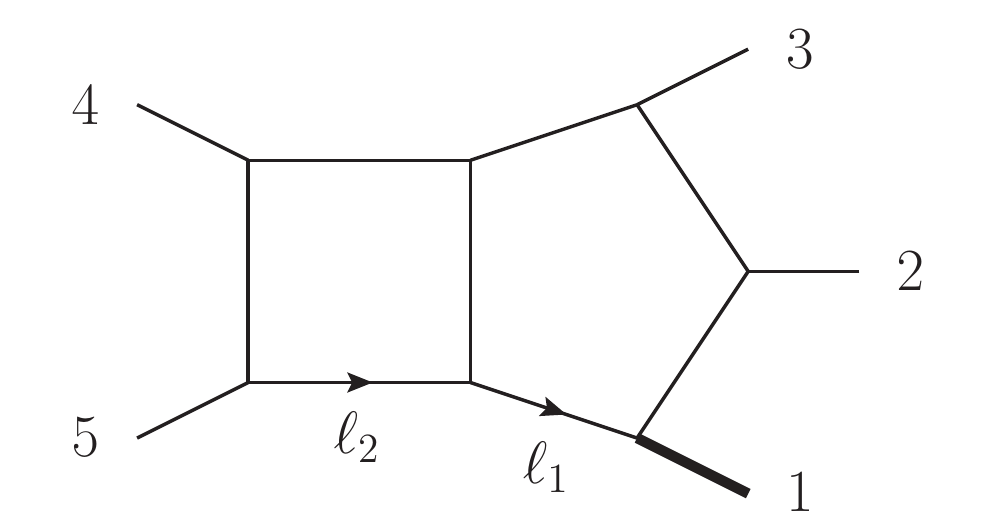}
    \caption{${\rm mzz}$}
    \label{fig:mzz}
\end{subfigure}
\begin{subfigure}{.5\textwidth}
    \centering
    \includegraphics[width=.8\textwidth]{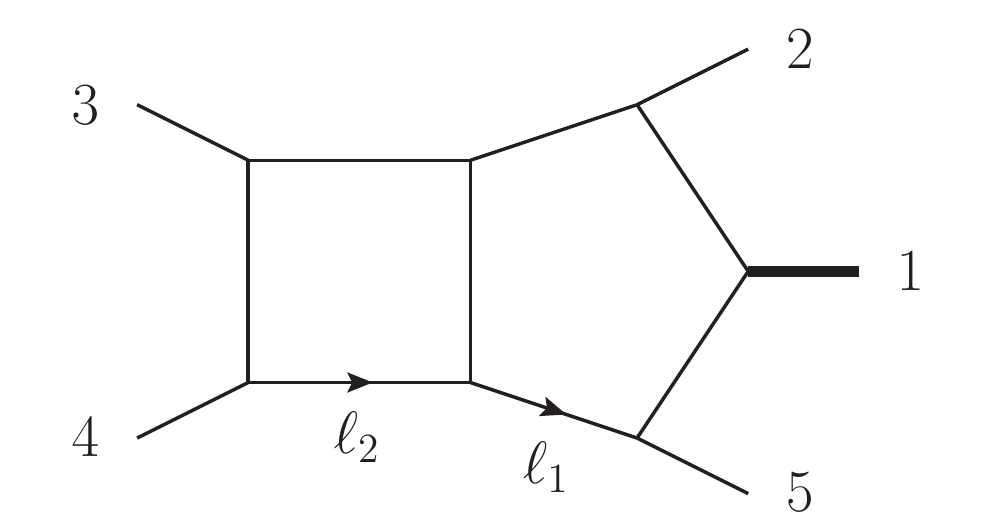}
    \caption{${\rm zmz}$}
    \label{fig:zmz}
\end{subfigure}%
\begin{subfigure}{.5\textwidth}
    \centering
    \includegraphics[width=.8\textwidth]{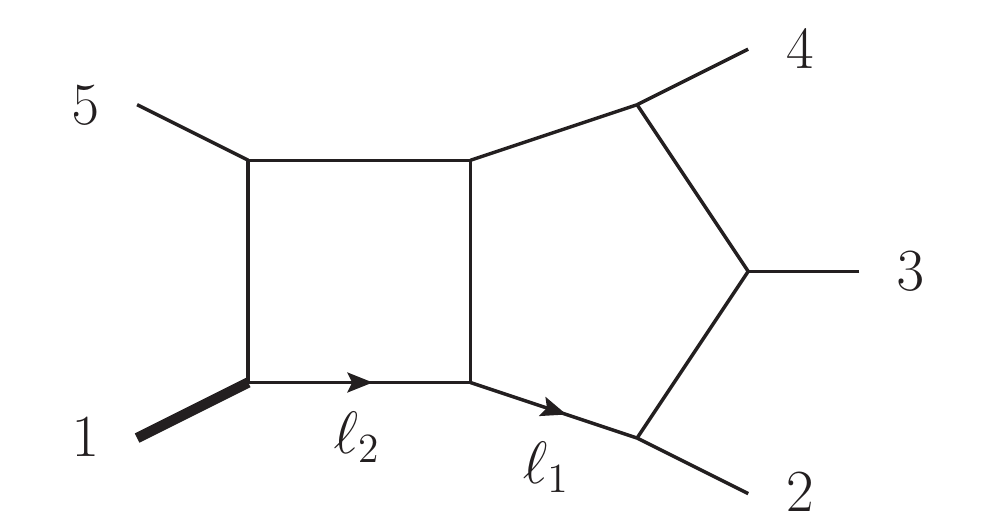}
    \caption{${\rm zzz}$}
    \label{fig:zzz}
\end{subfigure}
\caption[short]{Representative diagrams for the planar five-particle integral families with a single external massive leg at one and two loops. The arrows denote momentum flow. The thick line represents the massive external leg.}
\label{fig:families}
\end{figure}

The Feynman integrals appearing in planar five-particle amplitudes with a single massive external leg up to two-loop order belong to the families shown in figure~\ref{fig:families} and permutations thereof. We adopt the definitions of ref.~\cite{Abreu:2020jxa}. There is a single one-loop pentagon family, which we label by ${\rm 1L}$, and three two-loop pentabox families, which we label by ${\rm mzz}$, ${\rm zmz}$ and ${\rm zzz}$ depending on the location of the massive leg, as shown in figure~\ref{fig:families}.
We consider each family in the $S_4$ permutations of the external massless particles. 
The divergences in the integrals are regulated in dimensional regularization with $D = 4 - 2 \eps$.
The one-loop one-mass pentagon integrals are defined as
\begin{subequations} \label{eq:integral-famililies}
\begin{align} 
G_{{\rm 1L},\sigma}[\vec{\bf a}] = e^{\eps \gamma_E} \,\qty(\mu^2)^\epsilon\, \int \frac{\dd ^D\ell_1}{\ii \pi^{\frac{D}{2}}} \frac{1}{\prod\limits_{j=1}^{5} D^{a_j}_{{\rm 1L},\sigma,j}}\,, 
\end{align}
while the integrals of the two-loop one-mass pentabox families are given by
\begin{align}
G_{\tau,\sigma}[\vec{\bf a}] = e^{2 \eps \gamma_E} \,\qty(\mu^2)^{2\epsilon}\, \int \left(\prod_{i=1}^{2} \frac{\dd ^D\ell_i}{\ii \pi^{\frac{D}{2}}}\right) \frac{\prod\limits_{j=9}^{11} D^{-a_j}_{\tau,\sigma,j}}{\prod\limits_{j=1}^{8} D^{a_j}_{\tau,\sigma,j}}\, , \quad \tau \in \{ {\rm mzz} , {\rm zmz} , {\rm zzz}\} \,.
\end{align}
\end{subequations}
Here $\gamma_E$ is the Euler-Mascheroni constant, $\mu$ is the regularization scale, $\sigma$ is a permutation label, and a list of integers $\vec{\bf a}$ encodes the propagator exponents. 
In this paper we set $\mu=1$, which is equivalent to considering dimensionless ratios of the Mandelstam invariants in \cref{eq:Mand} to $\mu^2$.
The definitions of the inverse propagators $D_{\tau,\sigma}$ are gathered in \cref{tab:propagators}.
The last three propagators of the two-loop families are irreducible scalar products, i.e.\ auxiliary propagators introduced in order to express the numerators.
Therefore we assume that $\{ a_j \}_{j=9}^{11}$ are non-positive integers.

\begin{table}[t]
\centering
\begin{tabular}{cllll}
\toprule
$j$ & $D_{{\rm 1L},\sigma,j}$ & $D_{{\rm mzz},\sigma,j}$ & $D_{{\rm zmz},\sigma,j}$ & $D_{{\rm zzz},\sigma,j}$  \\
\midrule
1 & $(\ell_1)^2$ & $(\ell_1)^2$ & $(\ell_1)^2$ & $(\ell_1)^2$ \\
2 & $(\ell_1+p_{1})^2$ & $(\ell_1+p_{1})^2$ & $(\ell_1+p_{\sigma_5})^2$ & $(\ell_1+p_{\sigma_2})^2$ \\
3 &  $(\ell_1+p_{1}+p_{\sigma_2})^2$  & $(\ell_1+p_{1}+p_{\sigma_2})^2$ & $(\ell_1+p_{1}+p_{\sigma_5})^2$  & $(\ell_1+p_{\sigma_2}+p_{\sigma_3})^2$ \\
4 & $(\ell_1-p_{\sigma_4}-p_{\sigma_5})^2$ & $(\ell_1-p_{\sigma_4}-p_{\sigma_5})^2$ & $(\ell_1-p_{\sigma_3}-p_{\sigma_4})^2$  & $(\ell_1-p_{1}-p_{\sigma_5})^2$ \\
5 & $(\ell_1-p_{\sigma_5})^2$  & $(\ell_2)^2$ & $(\ell_2)^2$ & $(\ell_2)^2$ \\
6 &  & $(\ell_2-p_{\sigma_4}-p_{\sigma_5})^2$ & $(\ell_2-p_{\sigma_3}-p_{\sigma_4})^2$ & $(\ell_2-p_{1}-p_{\sigma_5})^2$ \\
7 &  & $(\ell_2-p_{\sigma_5})^2$ & $(\ell_2-p_{\sigma_4})^2$ &  $(\ell_2-p_{1})^2$ \\
8 &  & $(\ell_1-\ell_2)^2$ & $(\ell_1-\ell_2)^2$ & $(\ell_1-\ell_2)^2$ \\
9 &  & $(\ell_1-p_{\sigma_5})^2$ & $(\ell_2+p_{\sigma_5})^2$ & $(\ell_1-p_{1})^2$ \\
10 &  & $(\ell_2+p_{1})^2$ & $(\ell_2+p_{1}+p_{\sigma_5})^2$ & $(\ell_2+p_{\sigma_2})^2$ \\
11 &  & $(\ell_2+p_{1}+p_{\sigma_2})^2$ & $(\ell_1-p_{\sigma_4})^2$ & $(\ell_2+p_{\sigma_2}+p_{\sigma_3})^2$ \\
\bottomrule
\end{tabular}
\caption{Inverse propagators of the one-loop pentagon (${\rm 1L}$) and of the two-loop pentabox families (${\rm mzz}$, ${\rm zmz}$ and ${\rm zzz}$) for all permutations $\sigma=(\sigma_2\sigma_3\sigma_4\sigma_5)$ of the external massless legs. The representative diagrams are shown in figure~\ref{fig:families} for the standard ordering of the external legs, $\sigma_{\text{id}} = (2345)$.}
\label{tab:propagators}
\end{table}

\begin{figure}[t!]
\centering
\includegraphics[width=.4\textwidth]{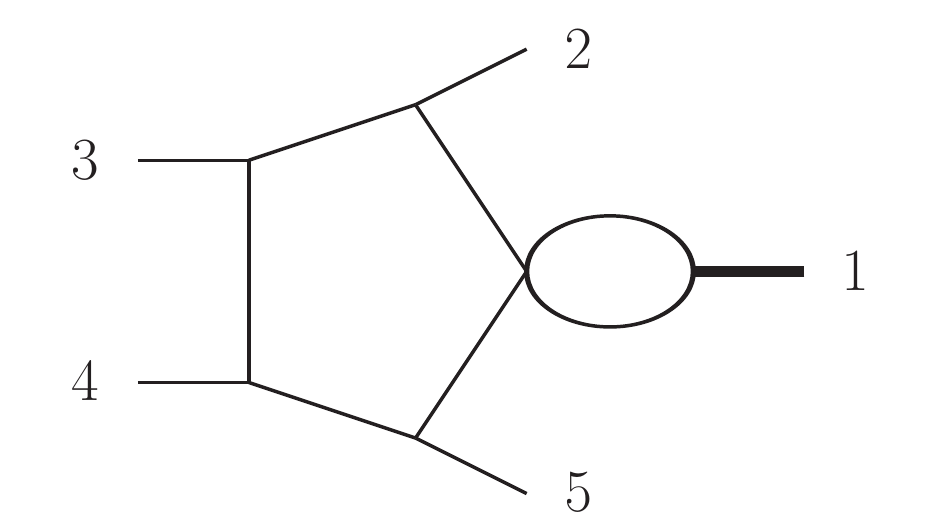}
\caption{One-loop squared integral family (${\rm 1L}^2$) appearing in the computation of two-loop five-particle amplitudes with a single external massive leg.}
\label{fig:1Lsquared}
\end{figure}

The two-loop scattering amplitudes also contain Feynman integrals which factorize into the product of two one-loop integrals of the pentagon family ${\rm 1L}$.
Some of these ``one-loop squared'' integrals are linearly independent from the two-loop pentaboxes.
They constitute an additional family shown in figure~\ref{fig:1Lsquared}, which we dub $\mathrm{1L}^2$.
One can express the integrals from this family in terms of transcendental functions by multiplying the expressions of the one-loop integrals which constitute them.
Since we are interested in constructing a set of \emph{algebraically} independent functions, we can omit the one-loop squared integrals from our considerations in \cref{sec:FunctionBasis}.

The linear span of all scalar integrals $G_{\tau,\sigma}[\vec{\bf a}]$ in a given family $\tau$ and orientation of the external legs $\sigma$, for any (allowed) set of indices $\vec{\bf a}$, has a finite-dimensional basis.
Any scalar integral $G_{\tau,\sigma}[\vec{\bf a}]$ can thus be expressed as a $\mathbb{Q}\left( \eps, X \right)$-linear combination of the master integrals, 
namely a linear combination with coefficients which are rational functions of the Mandelstam invariants and the dimensional regulator $\eps$.
This is typically achieved by solving linear systems of integration-by-parts identities (IBPs)~\cite{Tkachov:1981wb,Chetyrkin:1981qh,Laporta:2001dd}.
The choice of the master integrals is to a large extent arbitrary and can be exploited to simplify selected aspects of their computation.
In this paper, we take advantage of the integrals which satisfy the mathematical property known as transcendental purity \cite{Arkani-Hamed:2010pyv,Henn:2013pwa}.
We denote by $\vec{g}_{\tau,\sigma}$ the vector of pure master integrals for the family $\tau$ in the permutation $\sigma$.
The pure master integrals $\vec{g}_{\tau,\sigma}$ are expressed as linear combinations of scalar Feynman integrals,
\begin{align}
\vec{g}_{\tau,\sigma} = \sum_{\vec{\bf a}} \vec T_{\tau,\sigma}[\vec{\bf a}] \, G_{\tau,\sigma}[\vec{\bf a}] \,,
\end{align}
where the coefficients $\vec T_{\tau,\sigma}[\vec{\bf a}]$ are rational functions of the $X$ and $\eps$.  
The pure master integral bases for the families in figure~\ref{fig:families} have been constructed in ref.~\cite{Abreu:2020jxa} for the orientation $\sigma_{\text{id}}=\left(2345\right)$.
We obtain the other orientations by permutations according to~\cref{sigmaf}.

There are relations also among the integrals belonging to different families and in different permutations. 
We identified the relations in the $\mathbb{Q}$-linear spans
\begin{gather} \label{eq:integral-spans}
  \bigcup_{\sigma \in S_4} \, \vec{g}_{\mathrm{1L},\sigma} \,, \qquad\qquad  
  \bigcup_{{\footnotesize\begin{matrix}\tau \in \qty{\mathrm{mzz}, \mathrm{zmz}, \mathrm{zzz}, \mathrm{1L}^2 },\\\sigma \in S_4\end{matrix}}} \; \vec{g}_{\tau,\sigma} \,,
\end{gather}
of the one- and two-loop integrals respectively by analyzing their Symanzik polynomials (see e.g.\ ref.~\cite{Pak:2011xt}). 
The resulting numbers of linearly independent master integrals are shown in table~\ref{tab:MIsLinInep},
where for comparison we also show the number of master integrals in one orientation.
We observe that the dimensions of the linear spans in \cref{eq:integral-spans} can be likewise calculated by counting the number of nonequivalent scalar integral graphs, which suggests that all of the relations from the Symanzik-polynomial identifications follow from graph isomorphisms. We give the mappings among the pure master integrals in the ancillary file \texttt{master\_integral\_mappings.m}~\cite{supplemental}.

\begin{table}[ht]
\centering
\begin{tabular}{c*{2}{C{16ex}}}
\toprule
Loop order & Master integrals, $\sigma_\text{id}$ & Master integrals, $S_4$ \\
\midrule
$1^{\phantom{*}}$ & 13 &  56 \\
$2^{\phantom{*}}$ & 167 & 1361 \\
$2^*$ & 142 & 1244 \\
\bottomrule
\end{tabular}
\caption{
  Number of independent one- and two-loop master integrals in the unions of integral families $\qty{\mathrm{1L}}$ and $\qty{\mathrm{mzz}, \mathrm{zmz}, \mathrm{zzz}, \mathrm{1L}^2 }$ respectively.
  The second column corresponds to the integrals in the standard permutation $\sigma_\text{id}$, and the third column corresponds to the complete permutation orbit.
  The last row shows the number of non-factorizable two-loop master integrals in each case.}
\label{tab:MIsLinInep}
\end{table}

\subsection{Permutations and physical channels}
\label{sec:Permutations}

The master integrals of the families shown in \cref{fig:families,fig:1Lsquared} appear in the scattering amplitudes in various permutations of the external massless legs.
Following refs.~\cite{Badger:2019djh,Chicherin:2020oor,Badger:2021nhg}, we aim to express all permutations of the master integrals in terms of a common basis of special functions which are well-defined and can be evaluated numerically in a whole physical scattering region.
Without loss of generality we chose the latter to be the $s_{45}$ channel. 
We show that considering all $S_4$ permutations of the master-integral families allows us to relate any massive-particle production channel~\eqref{eq:process} to $45 \to 123$.
Let us consider a phase-space point $X$ of the physical channel $\sigma_4 \sigma_5 \to 1 \sigma_2 \sigma_3$ for some permutation $\sigma = \left(\sigma_2\sigma_3\sigma_4\sigma_5 \right) \in S_4$,
\begin{align}
X \in {\cal P}_{\sigma_4 \sigma_5} = \sigma \circ {\cal P}_{45} \,.
\end{align}
In order to evaluate the master integrals $\vec{g}_{\tau,\bar\sigma}$ of the family $\tau$ in the orientation $\bar\sigma$ at a point $X \in {\cal P}_{\sigma_4 \sigma_5}$,
we first map $X$ to a phase-space point $X'  \in {\cal P}_{45}$ using the inverse permutation,
\begin{align}
X' := \sigma^{-1} \circ X \in {\cal P}_{45} \,.
\end{align}
Then, we evaluate the same master integrals in the composed permutation $\sigma\bar{\sigma}$ at the corresponding point $X'$ in the $s_{45}$ channel, as shown by the following chain of equalities:
\begin{align} \label{eq:gsigmabar}
\vec{g}_{\tau, \bar\sigma}\left(X\right) = \vec{g}_{\tau, \bar\sigma} \left(\sigma \circ X' \right) = \vec{g}_{\tau, \sigma \bar\sigma} \left(X' \right) \,.
\end{align}

This is to be contrasted with an alternative approach often adopted in the literature,
which consists in considering the master integrals for a fixed ordering of the external legs and analytically continuing them to the other scattering regions.
The permutations of the master integrals in this case are performed only at the evaluation stage,
\begin{align} \label{eq:gsigmaX}
\vec{g}_{\tau, \sigma}\left(X\right) = \vec{g}_{\tau, \sigma_{\text{id}}}\left(\sigma \circ X\right) \,.
\end{align}
This however means that the integrals have to be evaluated a number of times equal to the number of required permutations for each desired phase-space point. Moreover, all the analytic simplifications which stem from cancellations among integrals in different  permutations of the external legs are missed. This not only makes the expressions for the amplitudes more obscure, but also prevents from subtracting the ultraviolet and infrared poles analytically, and introduces numerical cancellations which may affect the numerical stability of the evaluations.

\subsection{Canonical differential equations}
\label{sec:DEs}

The master integrals satisfy a system of first-order differential equations (DEs)~\cite{Kotikov:1990kg,Bern:1993kr,Remiddi:1997ny,Gehrmann:1999as}.
If the master integrals are pure, their DEs take a particularly simple canonical form \cite{Henn:2013pwa} where only logarithmic differential forms contribute.
In this work we employ the canonical DEs for $\vec g_{\tau,\sigma_{\text{id}}}$ that were constructed in ref.~\cite{Abreu:2020jxa},
\begin{equation} \label{eq:canonicalDEsSigma0}
\begin{aligned}
& \dd  \vec g_{\tau,\sigma_{\text{id}}}(X) = \eps \, \dd \tilde{A}_{\tau,\sigma_{\text{id}}}(X) \,  \vec g_{\tau,\sigma_{\text{id}}}(X) \,, \\
& \dd  \tilde{A}_{\tau,\sigma_{\text{id}}}(X)   = \sum_{i\in \mathbb{A}_{D_4}} a^{(i)}_{\tau,\sigma_{\text{id}}}\, \dd \log{W_i(X)} \,,
\end{aligned}
\end{equation}
where $a^{(i)}_{\tau,\sigma}$ are matrices of rational numbers, and $W_{i}$ are algebraic functions of the kinematic variables called \textit{letters} of the \textit{alphabet} $\mathbb{A}_{D_4}$.
Here the integrals are normalized such that their expansion around $\eps = 0$ starts with $\eps^0$,
\begin{align} \label{eq:geps}
\vec{g}_{\tau,\sigma}(X) = \sum_{w \geq 0}\eps^{w} \,\vec{g}^{\,(w)}_{\tau,\sigma} (X) \,,
\end{align}
so that their fourth order in $\eps$ is the highest required for the computation of two-loop scattering amplitudes up to order $\eps^0$.
The alphabet $\mathbb{A}_{D_4}$ is closed under the cyclic permutations of the external massless legs,
\begin{align} \label{eq:D4}
D_4 = \left\{ \left(2345\right), \left(5432\right) \right\} \,.
\end{align}

We are interested in a basis of transcendental functions sufficient to represent any planar two-loop five-point one-mass scattering amplitude through transcendental weight four.
We thus consider the master integrals in all orientations and construct an $S_4$-closure $\mathbb{A}_{S_4}$ of the cyclic alphabet $\mathbb{A}_{D_4}$ by
choosing a basis in the linear span of the permutation orbits $\sigma\circ\dd{\log{W_i}}$, for any $i\in \mathbb{A}_{D_4}$ and $\sigma \in S_4$.
We then straightforwardly generalize the canonical DEs~\eqref{eq:canonicalDEsSigma0} to all permutations $\sigma \in S_4$ of the integral families,
\begin{equation} \label{eq:canonicalDEs}
  \begin{aligned}
  & \dd  \vec g_{\tau,\sigma}(X) = \eps \, \dd  \tilde{A}_{\tau,\sigma}(X) \,  \vec g_{\tau,\sigma}(X) \,, \\
  & \dd  \tilde{A}_{\tau,\sigma}(X)   = \sum_{i \in \mathbb{A}_{S_4}} a^{(i)}_{\tau,\sigma}\, \dd \log{W_i(X)} \,.
  \end{aligned}
\end{equation}

The alphabet $\mathbb{A}_{S_4}$ is a subset of the non-planar alphabet recently established in ref.~\cite{Abreu:2021smk}, and for convenience we adopt the notation of the latter.
We summarize here the embedding of $\mathbb{A}_{D_4}$ and $\mathbb{A}_{S_4}$ into the non-planar alphabet of ref.~\cite{Abreu:2021smk}.
We explicitly spell out the expressions of the letters involved in our work in \cref{app:alphabet} and in the ancillary file \texttt{alphabet.m}~\cite{supplemental}.
The cyclic alphabet contains $58$ letters, which can be separated into two sets,
\begin{align}
\mathbb{A}_{D_4} = \mathbb{A}_{D_4}^{{\rm rel}}  \cup \mathbb{A}_{D_4}^{{\rm irrel}} \,,
\end{align}
of $49$ \textit{relevant} letters (those that appear in the solution of the DEs truncated at order $\eps^4$),
\begin{equation} \label{eq:AD4rel}
\begin{aligned}
\mathbb{A}_{D_4}^{{\rm rel}}  = \{\, & 1, 2, 5,\ldots, 16, 18, 19, 21, 22, 24, 25, 27, \ldots, 31, 33, 34, 45, 46, 57, 70, 93, \\
& 118, 120, 121, 123, 124, 126, 129, 130, \ldots, 134, 136, 137, 186, 195, 198 \} \,,
\end{aligned}
\end{equation}
and $9$ \textit{irrelevant} ones,
\begin{equation}
\mathbb{A}_{D_4}^{{\rm irrel}}  = \{32, 94, 117, 127, 135, 138, 161, 188, 197 \} \,.
\end{equation}
The permutation closure $\mathbb{A}_{S_4}$ contains $156$ letters, which again can be split into two sets,
\begin{align}
\mathbb{A}_{S_4} = \mathbb{A}_{S_4}^{{\rm rel}}  \cup \mathbb{A}_{S_4}^{{\rm irrel}} \,,
\end{align}
of $108$ relevant letters,
\begin{align} \label{eq:fullalphabet}
\mathbb{A}_{S_4}^{{\rm rel}}  = \{1, \ldots , 57, 70,\ldots, 93, 118, \ldots 137, 186, 187, 188, 195, \ldots, 198\} \,,
\end{align}
and $48$ irrelevant letters%
\footnote{
Let us note that $\mathbb{A}_{D_4}^{{\rm rel}} \subset \mathbb{A}_{S_4}^{{\rm rel}}$, but the irrelevant letters $\mathbb{A}_{D_4}^{{\rm irrel}}$ are distributed among both $\mathbb{A}_{S_4}^{{\rm rel}}$ and $\mathbb{A}_{S_4}^{{\rm irrel}}$.
},
\begin{align}
\mathbb{A}_{S_4}^{{\rm irrel}} = \{94, \ldots, 117, 138, \ldots, 161 \} \,.
\end{align}

\subsection{Solution of the DEs in terms of iterated integrals}
\label{sec:IteratedIntegrals}

The canonical DEs \cref{eq:canonicalDEsSigma0} can be solved straightforwardly order-by-order in the $\eps$-expansion~\eqref{eq:geps} in terms of Chen's iterated integrals~\cite{Chen:1977oja} (see also \cite{Brown:2013qva}) as
\begin{align} 
  \vec g^{\, (w)}_{\tau,\sigma} (X) = \sum_{w' = 0}^{w} \sum_{i_1,\ldots,i_{w'}\in\mathbb{A}_{S_4^{\mathrm{rel}}}} a^{(i_1)}_{\tau,\sigma} \cdot a^{(i_2)}_{\tau,\sigma}  \cdots  a^{(i_{w'})}_{\tau,\sigma} \cdot \vec{g}^{\, (w-w')}_{\tau,\sigma}(X_0) \ \left[W_{i_1},\ldots,W_{i_{w'}}\right]_{X_0}(X) \,, \label{eq:solDE}
\end{align}
where $\vec{g}^{\, (w)}_{\tau,\sigma}(X_0)$ are the values of the master integrals at an arbitrary initial point $X_0$, which we discuss in \cref{sec:InitialValues}.
The iterated integrals $\left[W_{i_1},\ldots,W_{i_{n}}\right]$ over $\dd{\log}$ kernels are defined as
\begin{equation} \label{eq:iterated_integral}
\begin{aligned} 
  \left[W_{i_1},\ldots,W_{i_{n}}\right]_{X_0}(X) &:= \int_{0}^{1} \dd t  \, \partial_t \log(W_{i_n}(\gamma(t)))\, \left[W_{i_1},\ldots,W_{i_{n-1}}\right]_{X_0}(\gamma(t)) \,, \\
  \left[\right]_{X_0}(X) &:= 1,
\end{aligned}
\end{equation}
where $\gamma$ is a path in the space of the kinematic variables connecting the initial point $\gamma(0) = X_0$ and $\gamma(1) = X$.
The number of iterations $n$ corresponds to the \textit{transcendental weight}. 
We consider the initial point $X_0$ as fixed and treat the iterated integrals as functions of the endpoint $X$.\footnote{
  The ($\mathbb{Q}$-linear combinations of) iterated integrals considered in this work are homotopy functionals,
  i.e.\ they are invariant under continuous deformations of the integration path $\gamma$. While this is in general not true for the separate iterated integrals shown in \cref{eq:iterated_integral}, it does hold for the combinations given in \cref{eq:solDE}, since the latter arise as the solution of DEs which satisfy integrability conditions. We assume that the same path $\gamma$ is used for all iterated integrals, and that it lies entirely in the analyticity region of the integrals $\mathcal{P}_{45}^+$. The considered iterated integrals therefore do not depend on the specific choice of $\gamma$.
}
The iterated integrals form a shuffle algebra with the product
\begin{align} \label{eq:shuffle_product}
\left[W_{a_1}, \ldots, W_{a_m} \right]_{X_0}(X) \left[W_{b_{1}}, \ldots, W_{b_{m}} \right]_{X_0}(X) = \sum_{\vec{c}\in\vec{a}\shuffle\vec{b}} \left[W_{c_1}, \ldots, W_{c_{m+n}} \right]_{X_0}(X) \,,
\end{align}
where $\vec{a}=\{a_1,\ldots,a_m\}$ and similarly for $\vec{b}$ and $\vec{c}$, and $\vec{a}\shuffle\vec{b}$ denotes all possible ways to shuffle $\vec{a}$ and $\vec{b}$.
The shuffle product allows us to linearize the relations between products of transcendental functions of different weights,
and therefore to systematically identify all functional relations through linear algebra.
This equips us with a powerful tool in the construction of a function basis which we discuss in \cref{sec:BasisConstruction}.

The components $\vec{g}^{\,(w)}_{\tau,\sigma}$ of the $\eps$-expansion of the pure master integrals are graded by the transcendental weight $w$.
This grading is manifest in \cref{eq:solDE} and is compatible with the shuffle product in \cref{eq:shuffle_product}.
In other words, $\vec{g}^{\,(w)}_{\tau,\sigma}$ and the iterated integrals can be assigned a $\mathbb{Z}_{\ge 0}$ charge.
Further gradings can be associated with changing signs of the square roots of the problem: $\sqrt{\Delta_5}$ (or equivalently $\trfive$) and $\sqrt{\Delta_3^{(i)}}$ with $i=1,2,3$.
The pure master integrals which are normalized by one of these square roots are \emph{odd} with respect to that square root, while the others are \emph{even}.
We can thus assign a $\mathbb{Z}_{2} = \{+,-\} = \{\text{even}, \text{odd}\}$ charge to each pure master integral for each of the four square roots.
One can easily verify that the $\dd{\log}$ forms in \cref{eq:solDE} can also be assigned the same charges (see appendix~\ref{app:alphabet}). 
Thus, it follows from \cref{eq:solDE,eq:shuffle_product} that the shuffle algebra of the iterated integrals with logarithmic integral kernels drawn from the alphabet $\mathbb{A}_{S_4}$ is graded by $\mathbb{Z}_{\ge 0} \times \left(\mathbb{Z}_{2}\right)^4$. This grading constitutes a useful organizing principle in the construction of the function basis discussed in \cref{sec:BasisConstruction}.

We stress that the choice of signs of the square roots is arbitrary and does not change within $\mathcal{P}_{45}^{+}$. In this paper we choose
\begin{equation} \label{eq:sqbranches}
 \sqrt{\Delta_3^{(i)}}>0\,, \qquad \Im\sqrt{\Delta_5} >0\,.
\end{equation}
The results for the opposite sign of each of the square roots can be obtained by flipping the sign of corresponding negatively-charged integrals.

\subsection{Initial values for the differential equations}
\label{sec:InitialValues}

In order to construct the DE solutions~\p{eq:solDE} up to transcendental weight $w$ we need to determine the initial values $\vec{g}^{\, (w)}_{\tau,\sigma}(X_0)$.
Since our aim is to find a function basis in the space of solutions, we have to take the algebraic relations between the initial values into account.
Our approach to this end is the following. 
We evaluate the master integrals at an initial point with sufficiently large accuracy,
and use the PSLQ algorithm~\cite{PSLQ} to identify integer relations among them.
This allows us to extract a generating set $\kappa$ of algebraically independent transcendental constants over $\mathbb{Q}$,
and to express the initial values $\vec{g}^{\, (w)}_{\tau,\sigma}(X_0)$ as graded polynomials $\mathbb{Q}\qty[\kappa]$.

We choose an initial point $X_0 \in \mathcal{P}_{45}^+$,
\begin{align} \label{eq:basepoint}
X_0 := \left(p_1^2=1\,,s_{12} = 3\,,s_{23} = 2\,,s_{34} = -2\,, s_{45} = 7\,, s_{15} = -2 \right) \,,
\end{align}
which satisfies the following requirements:
\begin{enumerate}

  \item $X_0$ introduces a minimal number of distinct prime factors.

  \item $X_0$ is invariant under the automorphisms of $\mathcal{P}_{45}$,
    \begin{align} \label{eq:automorphisms}
      \Sigma\left(\mathcal{P}_{45}\right) = \left\{ \left(2345\right),\, \left(2354\right),\, \left(3245\right),\, \left(3254\right)  \right\} \,.
    \end{align}
  In other words, it is invariant under the exchanges of momenta $2 \leftrightarrow 3$ and $4 \leftrightarrow 5$.
  
  \item The four letters which have indefinite sign inside $\mathcal{P}_{45}$ and depend linearly on the Mandelstam invariants vanish at $X_0$ (see \cref{app:alphpositivity}).
  
\end{enumerate}
In order to understand the first two requirements we recall that the master integrals considered in this work can be expressed in terms of multiple polylogarithms whose arguments are rational/algebraic functions of the external kinematics~\cite{Canko:2020ylt,Syrrakos:2020kba}. It is natural to expect that many of these arguments coincide if the phase-space point is degenerate. Choosing an initial point which is as symmetric as possible therefore reduces the number of distinct multiple polylogarithms in the initial values. Minimising the number of distinct prime factors in the initial point ---~and hence in the rational arguments~--- further simplifies the arguments of the polylogarithms evaluated there. These requirements therefore decrease the number of algebraically-independent initial values and thus facilitate the discovery of the relations through the PSLQ algorithm. We will make use of the last two properties in \cref{sec:explicit-repr,sec:Implementation}, where we construct an explicit representation of our function basis and design an algorithm for their numerical evaluation.

The next step consists in obtaining high-precision values of the master integrals at the initial point $X_0$~\eqref{eq:basepoint}.
We employ the analytic expressions in terms of Goncharov polylogarithms  (GPLs)~\cite{Goncharov:1998kja,Remiddi:1999ew,Goncharov:2001iea} constructed in refs.~\cite{Canko:2020ylt,Syrrakos:2020kba}. We use \texttt{GiNaC}~\cite{Bauer:2000cp,Vollinga:2004sn} to evaluate numerically the GPLs with $3000$-digit accuracy. We evaluate the permuted master integrals $\vec{g}_{\tau,\sigma}$ at $X= X_0$ by evaluating the expressions of refs.~\cite{Canko:2020ylt,Syrrakos:2020kba} for the standard ordering $\sigma_{\text{id}}$ of the external legs at permuted points, $X=\sigma \circ X_0$, as shown in \cref{eq:gsigmaX}.

The explicit GPL expressions of refs.~\cite{Canko:2020ylt,Syrrakos:2020kba} are given in an Euclidean region, while we are interested in the values of the Feynman integrals in the physical scattering regions. We perform the analytic continuation by adding a small positive imaginary part to the values of the independent Mandelstam invariants at the initial point $X_0$ as
\begin{multline} \label{eq:X0eta}
X_0^{(\eta)} := \\
\left( p_1^{2 \, (0)} + \ii c_1 \eta \,, s_{12}^{(0)} + \ii c_2 \eta \,, s_{23}^{(0)} + \ii c_3 \eta \,, s_{34}^{(0)} + \ii c_4 \eta \,, s_{45}^{(0)} + \ii c_5 \eta \,, s_{15}^{(0)} + \ii c_6 \eta \right) \,,
\end{multline}
where the superscript $(0)$ denotes the values at the initial point $X_0$~\eqref{eq:basepoint}, $\eta$ is an infinitesimal positive parameter, and the $c_i$'s are positive rational numbers. 
We choose the $c_i$'s randomly so that all scalar products $s_{ij}$ (including the non-adjacent ones) have a positive imaginary part
and all square roots in the alphabet are evaluated on the chosen branch (see \cref{eq:sqbranches}).

An additional complication we need to take care of is that some of the GPLs are divergent at $X_0$ (and permutations thereof).
Since the master integrals have no singularity in $\mathcal{P}_{45}^+$, these divergences are spurious and must cancel out.
We handle this by using the parameter $\eta$ in~\cref{eq:X0eta} as regulator and compute the limit $\eta \to 0^+$ of the divergent GPLs analytically using \texttt{PolyLogTools}~\cite{Duhr:2019tlz}.
The cancellation of divergences at all permutations of $X_0$ is a consistency check of our GPL-evaluation routine.
We further checked our results by evaluating all permutations of the master integrals also with the generalized power series expansion method~\cite{Moriello:2019yhu}. We used the initial values provided in ref.~\cite{Abreu:2020jxa} and integrated the canonical DEs with the \texttt{Mathematica} package \texttt{DiffExp}~\cite{Hidding:2020ytt} at $15$-digit accuracy. We evaluated the permuted master integrals by permuting the canonical DEs as discussed in section~\ref{sec:DEs}, rather than permuting the initial point as done for the GPLs. This way we verified also the consistency of the two approaches. We found full agreement.

\begin{table}[ht]
  \centering
  \begin{tabular}{cC{7ex}C{7ex}cc}
    \toprule
    & \multicolumn{2}{C{14ex}}{Linear span ($\oplus$ products)} & \multicolumn{2}{c}{Irreducible} \\
    \cmidrule(lr){2-3}\cmidrule(lr){4-5}
     Weight & $\Re$ & $\Im$ & $\Re$ & $\Im$  \\
    \midrule
    1 & 4\textsuperscript{$\ast$} & 1 & 4\textsuperscript{$\ast$} & 1\\
    2 & 12 & 4 & 5 & 0\\
    3 & 67 & 23  & 23 & 7 \\
    4 & 305 & 135 & 90 & 40\\
    \bottomrule
  \end{tabular}
  \caption{
    Number of independent transcendental constants at each weight.
    The second column shows the dimensions of the linear spans of constants together with (weight-graded) products of lower-weight constants.
    The third columns shows the number of irreducible constants.\\
    {\footnotesize $^\ast$ Only three real constants appear at weight one, but we add an extra constant to resolve reducible higher-weight constants.}
  }
  \label{tab:tc-dim}
\end{table}

We construct the generating set $\kappa^{(w)}$ at weight $w$ recursively starting from weight 0 where $\kappa^{(0)} \equiv \qty{1}$.
We consider the $\mathbb{Q}$-linear span $\mathcal{C}_0^{(w)}$ of the initial values at weight $w$, $\vec{g}^{\, (w)}_{\tau,\sigma}(X_0)$,
and weight-graded products of the lower-weight constants $\kappa^{(w^\prime)}$ with $w^\prime<w$.
We employ an adapted version of the program \verb|PSLQM3| from the \verb|MPFUN2020| package \cite{Bailey2021MPFUN2020AN}, which implements the three-level multipair PSLQ algorithm \cite{Bailey:1999nv,Bailey:2017}
to search for integer relations using the initial values evaluated to 2000 digits. We then confirm the identified relations at the 3000-digit precision level.
We then use these relations to construct a basis in $\mathcal{C}_0^{(w)}$, preferring the lower-weight products. The remaining basis elements of $\mathcal{C}_0^{(w)}$ are then deemed irreducible and assigned to $\kappa^{(w)}$.
Note that we require the generating set $\kappa$ to be real, while the initial values $\mathcal{C}_0^{(w)}$ are complex.
For the purpose of constructing the former, we view the complex initial values as $\mathbb{Z}_2 = \qty{\Re,\Im}$-graded algebra which induces the corresponding grading onto the polynomial ring $\mathbb{Q}[\kappa]$.
The number of linearly independent constants in $\mathcal{C}_0^{(w)}$ and the number of irreducible constants at each weight are shown in \cref{tab:tc-dim}.
The former characterizes the complexity of the integer relation finding problem.
In practice, we slightly reduce this complexity by looking for relations only within subspaces of $\mathcal{C}_0^{(w)}$ that are odd or even under the changes of square roots' signs (see \cref{sec:IteratedIntegrals}). We give the generating set $\kappa$ of transcendental constants in the ancillary file \texttt{transcendental\_constants.m}~\cite{supplemental}.

We conclude this section with some remarks.
First, we should stress that our numerical analysis cannot guarantee that the enumerated transcendental constants do not satisfy any additional algebraic relations over $\mathbb{Q}$.
However, as we explain in \cref{sec:BasisConstruction}, this would not jeopardize our function basis.
Second, we emphasize that, while finding the relations among the transcendental constants is essential for the construction of the function basis, we do not need to know the constants in the generating set analytically.
Furthermore, as we discuss in \cref{sec:BasisConstruction}, it is possible to absorb the transcendental constants into the definition of the basis functions such that
only $\ii \pi$ and $\zeta_3$ explicitly appear in the expressions for the master integrals in term of these functions.

\section{Function basis}
\label{sec:FunctionBasis}

In this section we present a basis of special functions in terms of which we can express all the planar two-loop five-particle amplitudes with one external off-shell leg up to order $\eps^0$. 
We dub these functions \textit{one-mass pentagon functions}, and we denote them by $\{f^{(w)}_i\}$, where $w$ is the transcendental weight and $i$ is a label.
We begin by discussing our strategy for constructing function bases using their iterated-integral representations in \cref{sec:BasisConstruction}.
Then, we present the expressions for the functions of the basis which are well suited for efficient numerical evaluation in \cref{sec:explicit-repr}.

\subsection{Construction of the basis using Chen's iterated integrals}
\label{sec:BasisConstruction}

The starting point are the canonical DEs~\eqref{eq:canonicalDEs} for the master integrals of the planar families shown in figure~\ref{fig:families}, each considered in all the $S_4$ permutations of the external massless legs.
We solve the DEs in terms of iterated integrals up to order $\eps^4$, or equivalently up to transcendental weight four, with the base point given by~\cref{eq:basepoint}.
The components of the $\epsilon$ expansion of the master integrals are not linearly independent, and the properties of the iterated integrals expose in a manifest way all the functional identities. 
Our aim is to identify a minimal set of linearly independent components, and to express any solution of the DEs as a polynomial in them. This minimal set constitutes the basis of one-mass pentagon functions, $\{f^{(w)}_i\}$. We refer the interested reader to ref.~\cite{Chicherin:2020oor} for a thorough discussion of this procedure, and give here just a brief outline.
Since the functional relations are uniform in the transcendental weight, we proceed recursively weight by weight.
At each weight $w$, we find a basis of the linear space spanned by the weight-$w$ master integral components over $\mathbb{Q}$. 
The dimensions of these bases are shown in the second column of \cref{tab:classificationcounting}.
Comparing to the third column in \cref{tab:MIsLinInep}, we observe that at each weight the number of linearly independent integrals is less than the one implied by the topological identifications.
Next, we use the shuffle algebra \eqref{eq:shuffle_product} obeyed by iterated integrals to remove the \textit{reducible} functions, i.e.\ those functions which can be expressed as products of lower-weight functions. Assuming that the lower-weight pentagon functions are already found,  $\{ f^{(w')}_i \}$ with $0<w'<w$, we form all their products having weight $w$, i.e.
\begin{align} \label{eq:product}
f_{i_1}^{(w_1)} f_{i_2}^{(w_2)} \ldots f_{i_k}^{(w_k)} 
\end{align}
such that $w_1+w_2+\ldots + w_k = w$, with $1\le w_i < w$ for $i=1,\ldots,k$ and $1\le k \le w$.
This way we can mod out the products from the weight-$w$ basis.
Repeating this procedure weight by weight up to weight four gives a basis algebraically independent functions $\{ f^{(w)}_i \}$, which we call the one-mass pentagon functions.
Their number is shown in the third column of table~\ref{tab:classificationcounting}. 

The iterated-integral representation of the master integrals in \cref{eq:solDE} involves the initial values $\vec{g}^{(w)}_{\tau,\sigma}(X_0)$.
Thus, for our strategy to succeed, the relations between these transcendental constants have to be taken into account.
We identified these relations in \cref{sec:InitialValues} with a PSLQ-based analysis.
As a result, it is in principle possible that certain algebraic relations were overlooked.
Even if it were the case, the pentagon function classification of this section would not be affected.
Indeed, the linear and algebraic independence of the functions persists upon the symbol projection~\cite{Goncharov:2010jf,Duhr:2011zq,Duhr:2012fh},
which is obtained by putting to zero all the non-rational constants in the iterated integral representation.
In other words, the number of pentagon functions coincides with the number of linearly independent and irreducible symbols, and is therefore minimal.

\begin{table}[ht]
  \centering
  \begin{tabular}{c *3{C{13ex}}}
    \toprule
    Weight & Linearly independent & Irreducible & Irreducible, cyclic \\
    \midrule
    1 & 11 & 11 & 6 \\
    2 & 86 & 25 & 8 \\
    3 & 483 & 145 & 31 \\
    4 & 1187 & 675 & 113\\
    \bottomrule
  \end{tabular}
  \caption{
    Counting of the master integral components at each transcendental weight. The second column corresponds to the number of the $\mathbb{Q}$-linearly independent components.
    The third column shows the number of irreducible components, i.e.\ the number of functions in the basis. The fourth column shows the number of functions sufficient to express the master integrals in the cyclic permutations $D_4$~\eqref{eq:D4} only.
}
\label{tab:classificationcounting}
\end{table}

The one-mass pentagon functions inherit the $\mathbb{Z}_{\ge 0}\times \left(\mathbb{Z}_2\right)^4$ grading from the pure master integrals (see section~\ref{sec:IteratedIntegrals}).
Since the pure master integral normalizations involve at most one square root, the sets of functions with negative charge with respect to each square root are disjoint.
The number of basis functions with negative $\mathbb{Z}_2$ charge with respect to each of the square roots is shown in table~\ref{tab:charges}.
The charges of all one-mass pentagon functions are provided in the ancillary file \texttt{pfuncs\_charges.m}~\cite{supplemental}. 
\begin{table}[h!]
\begin{center}
\begin{tabular}{ccccc}
\toprule
Weight & $\trfive$ & $\sqrt{\Delta_3^{(1)}}$ & $\sqrt{\Delta_3^{(2)}}$ & $\sqrt{\Delta_3^{(3)}}$ \\ 
\midrule
1 & 0 & 0 & 0 & 0 \\
2 & 0 & 1 (1) & 1 (0) & 1 (0) \\
3 & 12 (1) & 5 (3) & 5 (0) & 5 (0) \\
4 & 87 (8) & 18 (9) & 18 (0) & 18 (1) \\
\bottomrule
\end{tabular}
\end{center}
\caption{Number of one-mass pentagon-functions with negative $\mathbb{Z}_2$ charge with respect to each of the square roots. In parentheses are the corresponding numbers for the cyclic subset specified in~\cref{eq:cyclpf}.}
\label{tab:charges}
\end{table}

The procedure described above to construct a basis out of the pure master integral components leaves a lot of freedom in choosing the basis elements $\{f^{(w)}_i\}$. We profit from this freedom to construct a basis which fulfills several additional constraints. 

First, it is convenient to reduce as much as possible the amount of transcendental constants in the $\eps$-components of the pure master integrals written as polynomials over $\mathbb{Q}$ of the pentagon functions.
We absorb the majority of those constants in the definition of the one-mass pentagon functions. The only leftover transcendental constants are powers of $\ii \pi$ and $\zeta_3$. 
For example, at weight two we have
\begin{align}
g^{(2)}_{\tau,\sigma} \in & \left\langle \{f^{(2)}_i\}_{i=1}^{25}, \{ f^{(1)}_i f^{(1)}_j \}_{i,j=1}^{11} , \{ \ii  \pi f^{(1)}_{i}\}_{i=1}^{11} , \pi^2  \right\rangle_{\mathbb{Q}} \,,
\end{align}
where by $\langle A  \rangle_{\mathbb{Q}}$ we denote the linear span of the set of pentagon functions and constants $A$ over $\mathbb{Q}$.
As a consequence, and in contrast with the basis of functions defined for the massless case in ref.~\cite{Chicherin:2020oor},
no constant in our basis has a non-trivial charge with respect to changing the signs of the square roots. 
We observe that, while the weight-four components of the master integrals contain products of all weight-one and two pentagon functions, only a subset of the weight-three ones ---~$\{f^{(3)}_i\}_{i = 4}^{120}$~--- appear.

The second useful constraint is the separation of the minimal subset of functions relevant for the cyclic permutations of the integral families.
If one is interested in fully color-ordered scattering amplitudes, it suffices to consider only the permutations of the master integrals which preserve the cyclic ordering of the particles, namely $\sigma \in D_4$~\eqref{eq:D4}.
It is therefore meaningful to separate a minimal subset of the one-mass pentagon functions which contribute in the two permutations $D_4$. We call them {\em cyclic} pentagon functions. They are counted in the fourth column of table~\ref{tab:classificationcounting}. Only the 49 relevant letters of the cyclic alphabet $\mathbb{A}_{D_4}^{{\rm rel}}$~\p{eq:AD4rel} contribute in the iterated integral expressions of these master integrals.
Explicitly, the cyclic pentagon functions are
\begin{align}
\left\{ f^{(1)}_i \right\}_{i=1}^{6} \cup \left\{ f^{(2)}_i \right\}_{i=3,\ldots,7,16,17,23}\cup \left\{ f^{(3)}_i \right\}_{i=1}^{31} \cup \left\{ f^{(4)}_i \right\}_{i=1}^{113} \,. \label{eq:cyclpf}
\end{align}
The expressions of the master integrals $\vec{g}_{\tau,\sigma}$ in the cyclic permutations $\sigma \in D_4$ thus involve only these functions. 

The third property we made manifest stems from the observations made in the computations of the two-loop amplitudes in
refs.~\cite{Badger:2021nhg,Badger:2021ega,Abreu:2021asb}.
Of the 49 letters of the cyclic alphabet $\mathbb{A}_{D_4}^{{\rm rel}}$~\p{eq:AD4rel}, the following six, 
\begin{align}
\mathcal{Z} := \{ W_{18},\,W_{25},\,W_{34},\,W_{45},\,W_{46},\,W_{57} \} \,, \label{eq:Zlett}
\end{align}
drop out from the amplitudes truncated at order $\eps^0$, despite being present in the expressions of the two-loop master integrals. In the same works it was also observed that the letter $W_{198}=\sqrt{\Delta_5}$, which is present in the amplitudes, drops out from the properly defined finite remainders. This is reminiscent of the behavior of the letter $\sqrt{\Delta_5}$ of the massless pentagon alphabet~\cite{Chicherin:2017dob}, which has been observed to drop out from several massless two-loop five-particle finite remainders. This phenomenon is expected to be a general feature, and the first steps towards an explanation have been taken in the context of cluster algebras~\cite{Chicherin:2020umh}.
Based on the experience with the massless scattering, it is thus natural to conjecture that $W_{198}$ drops out of the finite remainders for any one-mass five-particle scattering process up to two loops.
To make this explicit, we further refine the cyclic pentagon functions~\p{eq:cyclpf} by confining the letters $\mathcal{Z}\cup\{\sqrt{\Delta_5}\}$ to a minimal subset of them. This way, expressing the known two-loop five-particle amplitudes with one external massive leg~\cite{Badger:2021nhg,Badger:2021ega,Abreu:2021asb} in terms of our cyclic pentagon function basis~\p{eq:cyclpf}, we avoid the spurious appearance of the letters $\mathcal{Z}$ altogether.
Similarly, the finite remainders are free from the pentagon functions involving the letter $\sqrt{\Delta_5}$.
In addition to analytic simplifications, this has the benefit of making the numerical evaluation more efficient, as we avoid evaluating functions involving spurious integration kernels which would otherwise lead to numerical cancellations.

We provide the expressions of the independent master integrals in terms of one-mass pentagon functions in the ancillary file \texttt{mi2pfuncs.m}~\cite{supplemental}. The other master integrals can be obtained through the mappings in \texttt{master\_integral\_mappings.m} (see \cref{sec:IntegralFamilies}). We spell out the one-mass pentagon functions in terms of Chen's iterated integrals and transcendental constants in \texttt{pfuncs\_iterated\_integrals.m}. 

The one-mass pentagon functions are by construction closed under permutations of the external massless legs. In other words, any permutation $\sigma \in S_4$ of a pentagon function can be expressed in terms of (graded) polynomials in the pentagon functions,
\begin{align}
\left(\sigma \circ f^{(w)}_i\right)(X) \in \mathbb{Q}\left[ \left\{ f^{(w')}_j(X) \right\}_{w',j} \right] \,.
\end{align}
This allows to evaluate the pentagon functions in any physical $2\to3$ scattering region with production of a massive particle following the same procedure discussed in \cref{sec:Permutations} for the master integrals.
We provide all $S_4$ permutations of the one-mass pentagon functions in the folder \texttt{pfuncs\_permutations/} of the ancillary files~\cite{supplemental}.

\subsection{Explicit representations}
\label{sec:explicit-repr}

In the previous section we discussed how we identified a minimal set of irreducible iterated integrals which is sufficient to write down the $\eps$-expansion of any pure master integral up to transcendental weight four.
For this purpose it was crucial to express the master integrals, and thus the functions in the basis, in terms of Chen's iterated integrals.
In this section we present expressions for the basis functions which are well suited for their numerical evaluation. We give them in the ancillary file \texttt{pfuncs\_expressions.m}~\cite{supplemental}.
Following refs.~\cite{Gehrmann:2018yef,Chicherin:2020oor} we provide expressions in terms of logarithms and dilogarithms up to transcendental weight two,
and one-fold integral representations for the weight-three and four functions, as first suggested in ref.~\cite{Caron-Huot:2014lda}.
We constructed the expressions in terms of logarithms and dilogarithms using the strategy proposed in ref.~\cite{Duhr:2011zq} (see also ref.~\cite{Chicherin:2020oor}).

\subsubsection{Weight-one pentagon functions}
\label{sec:explicit-repr:w1}

We choose the 11 weight-one pentagon functions as follows,
\begin{equation}
\begin{aligned}
f^{(1)}_1 & = \log (p_1^2) 
\,,\quad
& f^{(1)}_2 &= \log(- s_{34}) 
\,,\quad
&f^{(1)}_3 & = \log(s_{12}) 
\,,\quad \\ 
f^{(1)}_4 & = \log(- s_{15}) 
\,,\quad
&f^{(1)}_5 & = \log(s_{23}) 
\,,\quad
&f^{(1)}_6 & = \log(s_{45}) 
\,,\quad \\
f^{(1)}_7 & = \log(- s_{25}) 
\,,\quad
&f^{(1)}_8 & = \log(- s_{24}) 
\,,\quad
&f^{(1)}_9 & = \log(- s_{35}) 
\,,\quad \\
f^{(1)}_{10} & = \log(s_{13}) 
\,,\quad
&f^{(1)}_{11} &= \log(- s_{14})\,. &
\end{aligned}
\end{equation}
It is easy to see that they are well-defined in the physical region ${\cal P}_{45}$~\p{eq:P45}, namely they have no branch cut within it. 
As mentioned in section~\ref{sec:BasisConstruction}, all the transcendental constants which depend on the arbitrary base point~\eqref{eq:basepoint} are absorbed inside the iterated integral representation, e.g.\ 
\begin{align}
f_2^{(1)}(X) = \log(- s_{34}) = [W_{9}]_{X_0} + \log(2)\,,
\end{align}
such that only $\ii  \pi$ appears explicitly at weight one in the expressions of the pure master integrals in terms of pentagon functions.

\subsubsection{Weight-two pentagon functions}
\label{sec:explicit-repr:w2}

From the analysis of the iterated integral solution of the canonical DEs for the master integrals we have identified 25 irreducible weight-2 functions. 
In order to present them in a more compact way, we introduce the following short-hand notations,
\begin{equation}
\begin{aligned}
& L(a,b):= {\rm Li}_2 \left( 1- \frac{a}{b}\right) \,,\\
& M(a,b):= {\rm Li}_2 \left( \frac{a}{b} \right)  + \log\left(- \frac{a}{b}\right)\log\left(1-\frac{a}{b}\right) + \ii  \pi \log\left(1- \frac{a}{b}\right) \,.
\end{aligned}
\end{equation}
The functions $L(a,b)$ and $M(a,b)$ are well-defined and single-valued for $a,b>0$ and for $a<0,\,b>0$, respectively. We choose the first 16 weight-two functions so that their iterated integral representation involves only alphabet letters which are linear in the Mandelstam invariants,
\begin{equation}
\begin{aligned}
f^{(2)}_1 &= L(s_{13},p_1^2)\,,\quad & 
f^{(2)}_2 &= M(s_{14},p_1^2)\,,\quad &
f^{(2)}_3 & = L( s_{12},p_1^2)\,, \quad \\ 
f^{(2)}_4 &= M(s_{15},p_1^2)\,, \quad &
f^{(2)}_5 &= L(s_{12},s_{45})\,, \quad &
f^{(2)}_6 & = M(s_{34},s_{12})\,,\quad \\
f^{(2)}_{7} &= M(s_{15},s_{23})\,,\quad &
f^{(2)}_8 &= L(s_{13},s_{45})\,, \quad &
f^{(2)}_{9} &= M(s_{35},s_{12})\,, \quad \\
f^{(2)}_{10} &= M(s_{25},s_{13})\,, \quad &
f^{(2)}_{11} &= M(s_{24},s_{13})\,, \quad &
f^{(2)}_{12} &= M(s_{14},s_{23})\,,\quad \\
f^{(2)}_{13} &= L(s_{24},s_{15})\,, \quad &
f^{(2)}_{14} &= L(s_{25},s_{14})\,,\quad &
f^{(2)}_{15} &= L(s_{35},s_{14})\,,\quad \\
f^{(2)}_{16} &= L(s_{34},s_{15})\,. \quad & \\
\end{aligned}
\end{equation}
All these functions are well-defined in the physical scattering region ${\cal P}_{45}$~\p{eq:P45}.

Then we have 6 weight-two functions whose iterated integral representation also involves alphabet letters which are quadratic in the Mandelstam invariants,
\begin{equation}
\begin{aligned}
f^{(2)}_{17} & = {\rm Li}_2 \left( 1-\frac{s_{12}s_{15}}{p_1^2 s_{34}}\right)   \;,\; &
f^{(2)}_{18} = {\rm Li}_2 \left( 1-\frac{s_{12}s_{14}}{p_1^2 s_{35}}\right)   \;, \\
f^{(2)}_{19} &= {\rm Li}_2 \left( 1-\frac{s_{13}s_{15}}{p_1^2 s_{24}}\right)   \;,\; &
f^{(2)}_{20} = {\rm Li}_2 \left( 1-\frac{s_{13}s_{14}}{p_1^2 s_{25}}\right)   \;, \\
f^{(2)}_{21} & = {\rm Li}_2 \left( 1-\frac{s_{12}s_{13}}{p_1^2 s_{45}}\right)   \;,\; &
f^{(2)}_{22} = {\rm Li}_2 \left( 1-\frac{s_{14}s_{15}}{p_1^2 s_{23}}\right) - 2 \ii  \pi \log \left( \frac{s_{14}s_{15}}{p_1^2 s_{23}} -1\right) \,.
\end{aligned}
\end{equation}
The first 5 of these functions are well-defined in the physical region ${\cal P}_{45}$, since the dilogarithm arguments cannot cross the branch cut within ${\cal P}_{45}$. In order to see that $f^{(2)}_{22}$ is well-defined as well we need to verify that the argument of the logarithm is positive inside ${\cal P}_{45}$. Indeed we find that $s_{14} s_{15} - p_1^2 s_{23} = W_{33}$, which in appendix~\ref{app:alphpositivity} we prove to be positive in $\mathcal{P}_{45}$.

We choose the remaining 3 weight-two functions to be given by the so-called three-mass triangles, i.e.\ they correspond (up to an algebraic prefactor) to the finite part of the one-loop triangle integrals with three-external off-shell legs (see e.g.\ ref.~\cite{Chavez:2012kn}). In order to write them down explicitly, we introduce the three-mass-triangle function, 
\begin{equation}  \label{eq:trifun}
\begin{aligned}
& {\rm Tri}^{(\rho)}(a,b,c) := 2 \, {\rm Li}_2 \left( 1 - \frac{2 a}{a-b+c - \sqrt{\lambda(a,b,c)}}\right) + 2 \, {\rm Li}_2 \left( 1 - \frac{2 a}{a+b-c - \sqrt{\lambda(a,b,c)}}\right) \\ 
& \ \ + \frac{1}{2} \log^2 \left( -1+\frac{2a}{a+b-c+ \sqrt{\lambda(a,b,c)}} \right) +
\frac{1}{2} \log^2 \left( -\rho +\frac{2a \rho}{a+b-c- \sqrt{\lambda(a,b,c)}} \right) \\
& \ \ + \frac{1}{2} \log^2 \left( \frac{a-b+c-\sqrt{\lambda(a,b,c)}}{a+b-c- \sqrt{\lambda(a,b,c)}} \right) -   \frac{1}{2} \log^2 \left( \rho \, \frac{a-b+c+\sqrt{\lambda(a,b,c)}}{a+b-c+ \sqrt{\lambda(a,b,c)}} \right)  + \frac{\pi^2}{3} \\ 
& \ \ + \ii \pi \delta_{\rho,-1} \log \left( \frac{a-b+c+ \sqrt{\lambda(a,b,c)}}{-a+b-c+ \sqrt{\lambda(a,b,c)}}\right) + \ii \pi \delta_{\rho,-1} \log \left( \frac{-a+b+c+ \sqrt{\lambda(a,b,c)}}{a-b-c+ \sqrt{\lambda(a,b,c)}}\right) \,,
\end{aligned}
\end{equation}
where $\rho = \pm 1$ and the K\"allen function $\lambda$ is defined in~\cref{eq:Kaellen}. In terms of this auxiliary function the last $3$ weight-two pentagon functions are given by 
\begin{equation} \label{eq:tri-w2}
\begin{aligned}
& f^{(2)}_{23} = {\rm Tri}^{(+1)}(s_{45},s_{23},p_1^2) \,,\\
& f^{(2)}_{24} = {\rm Tri}^{(-1)}(s_{34},s_{25},p_1^2) \,, \\
& f^{(2)}_{25} = {\rm Tri}^{(-1)}(s_{24},s_{35},p_1^2) \,. 
\end{aligned}
\end{equation}
Each of these three functions has a negative charge with respect to changing the sign of the three-mass triangle square root $\sqrt{\Delta^{(i)}_3}$ with the same arguments in~\cref{Deltalambda}. Indeed, these are the three functions with non-trivial charge at transcendental weight two in table~\ref{tab:charges}. In appendix~\ref{app:alphpositivity} we show that the functions in~\cref{eq:tri-w2} are well-defined in ${\cal P}_{45}$.

Finally, let us recall that 8 weight-two pentagon functions, $\{ f^{(2)}_i \}_{i=3,\ldots,7,16,17,23}$, belong to the cyclic subset listed in~\cref{eq:cyclpf}. In other words, only this subset appears in the permutations $D_4$ of the master integrals. These weight-two functions do not involve the letters in $\mathcal{Z}$~\p{eq:Zlett} nor $W_{198} = \sqrt{\Delta_5}$.

\subsubsection{Weight-three pentagon functions}
\label{sec:explicit-repr:w3}

At transcendental weights three and four we construct a representation of the pentagon functions in terms of one-dimensional integrals, which can be efficiently evaluated numerically~\cite{Chicherin:2020oor}. The starting point is their expression in terms of Chen's iterated integrals, which for the 145 weight-three functions is given schematically by
\begin{align}
f^{(3)}_i (X) & = \sum_{j_1,j_2,j_3} \tau^{(0)}_{i,j_1,j_2,j_3} 
[ W_{j_1}, W_{j_2}, W_{j_3} ]_{X_0}(X)  + \sum_{j_1,j_2} \tau^{(1)}_{i,j_1,j_2}  [W_{j_1},W_{j_2}]_{X_0}(X) \notag\\ & + \sum_{j_1} \tau^{(2)}_{i,j_1} [W_{j_1}]_{X_0}(X) +  \tau^{(3)}_i \,, \label{eq:3foldint}
\end{align}
where $\tau^{(w)}_{I}$ for any set of indices $I$ denotes weight-$w$ transcendental constants, and $\tau^{(0)}_I$ are rational numbers. It is worth noting that the iterated integrals do not involve the letter $W_{198} = \sqrt{\Delta_5}$, and the letters in the set $\mathcal{Z}$~\p{eq:Zlett} are absent in the cyclic pentagon functions, $\{ f^{(3)}_i \}_{i=1}^{31}$~\p{eq:cyclpf}. The criteria discussed in section~\ref{sec:BasisConstruction} therefore do not forbid any weight-three pentagon function to appear in the amplitudes and finite remainders.
Invoking the weight-one and two pentagon functions defined previously we can implement the two inner integrations in the three-fold integral in the first term on the right-hand side of~\cref{eq:3foldint}, and the inner integration in the second term. As a result, the weight-three pentagon functions take the following one-fold integral form, 
\begin{align}
f^{(3)}_i (X) = \sum_{j,k} c_{i,j,k}\int^1_0 \dd t \,\partial_t \log \left( W_j(t) \right) \, h^{(2)}_{k}(t)  + \tau^{(3)}_i \,, \label{eq:1foldw3}
\end{align} 
where $c_{i,j,k}$ are rational numbers, $\tau^{(3)}_i$ are transcendental weight-three constants, and $h^{(2)}_k (X)$ are weight-two functions. The latter are polynomial in $\ii \pi$ and the weight-one and two pentagon functions,
\begin{align}
h^{(2)}_k \in \left\langle \{ f^{(2)}_i \}_{i=1}^{25}, \{ f^{(1)}_i f^{(1)}_j \}_{i,j=1}^{11} , \{ \ii \pi f^{(1)}_j \}_{j=1}^{11} , \pi^2 \right\rangle_{\mathbb{Q}} \,.
\end{align}
The integration in~\cref{eq:1foldw3} is pulled back to the interval $[0,1]$ by an arbitrary contour $\gamma$, which connects the base point $X_0=\gamma(0)$ to an arbitrary point $X=\gamma(1) \in {\cal P}_{45}^+$. To simplify the expression we used the short-hand notations 
\begin{align}
h_k^{(2)}(t) := h_k^{(2)}\left(X=\gamma(t)\right) \,, \qquad \qquad W_j(t) := W_j\left(X=\gamma(t)\right) \,.
\end{align}
The construction of the explicit contour $\gamma$ is discussed in detail in section~\ref{sec:Implementation}. For the purposes of this section it suffices to know that it lies entirely within the physical region ${\cal P}_{45}^+$.
For the sake of completeness, let us note that not all 108 letters~\p{eq:fullalphabet} appear in the logarithmic integration kernel in~\cref{eq:1foldw3}: the letter $W_{198} = \sqrt{\Delta_5}$ and 24 of the 54 letters which are quadratic in the Mandelstam invariants are absent.

Prior to attempting the numerical evaluation of the one-dimensional integrations in \cref{eq:1foldw3} we need to verify that the integrations are well-defined for a path inside the physical region ${\cal P}_{45}^+$. The functions $h_k^{(2)}(X)$ are polynomials in the pentagon functions of weights one and two, and thus they do not have singularities nor branch cuts anywhere in ${\cal P}_{45}^+$. We only need to inquire about possible singularities of the algebraic functions $\partial_t \log\left( W_j(t)\right)$, which are located at $W_j(X) = 0$. In appendix~\ref{app:alphpositivity} we show that $20$ letters can vanish inside ${\cal P}_{45}^+$. We call them {\em spurious letters}. There are $4$ spurious letters which are linear in the Mandelstam invariants,
\begin{align} \label{eq:spurious_linear}
{\cal S}_{\ell} = \{ & W_{16},\,W_{17},\, W_{19},\,W_{20} \} \,,
\end{align} 
while 16 are quadratic,
\begin{equation}
\begin{aligned}
{\cal S}_q = \{ & W_{35},\,W_{36},\,W_{38},\,W_{39},\,W_{40},\,W_{41},\,W_{43},\,W_{44},\,\\
& W_{72},\,W_{74},\,W_{78},\,W_{80},\,W_{82},\,W_{84},\,W_{88},\,W_{90} \} \,.
\end{aligned}
\end{equation} 
We denote the set of all spurious letters by $\mathcal{S}$,
\begin{align}
\mathcal{S} = {\cal S}_{\ell} \cup {\cal S}_q \,.
\end{align}
While the linear spurious letters ${\cal S}_{\ell}$ vanish at the base point $X_0= \gamma(0)$,
\begin{align}
W \Bigr|_{X=X_0} = 0   \qquad \forall \; W \in {\cal S}_{\ell}\,,
\end{align}
none of the quadratic spurious letters ${\cal S}_{q}$ vanishes at the base point.
Moreover, the weight-two functions $h^{(2)}_k(X)$ which in~\cref{eq:1foldw3} accompany the logarithms of the spurious letters, $\log W_j(t)$ with $W_j \in {\cal S}$, vanish on the locus where the corresponding letter vanishes, $\{ X: W_j(X) = 0 \}$.
The latter is straightforward to verify since the $h^{(2)}_k(X)$ which accompany the $\dd \log$s of spurious letters in the expressions of the weight-three functions do not involve the complicated three-mass-triangle pentagon functions.
We can thus conclude that the logarithmic integration kernels in the weight-three one-fold integrals~\p{eq:1foldw3} may have at most a simple pole singularity,
\begin{align} \label{eq:hvanish1}
\partial_t \log(W_j(t)) = {\cal O}\left(\frac{1}{t-t_0}\right) \,,
\end{align} 
in the neighbourhood of $t=t_0$ such that $W_j(t_0) = 0$, and that such simple poles are suppressed by the vanishing of the corresponding weight-two functions $h^{(2)}_k$,
\begin{align} \label{eq:hvanish2}
h^{(2)}_k(t) = {\cal O}(t-t_0) \,.
\end{align}
For the linear spurious letters $W_j \in {\cal S}_{\ell}$ the spurious singularity is at the base point $t_0 = 0$, while the quadratic spurious letters $W_j \in {\cal S}_{q}$ produce spurious singularities for $t_0 \in(0,1]$.
It is convenient to confine the weight-three functions with spurious singularities in a minimal set of 2 cyclic and 14 non-cyclic pentagon functions. This classification can be summarised schematically as
\begin{align}
\left\{ f^{(3)}_{i}\right\}_{i=1}^{145} \; : \qquad
\overbrace{1,\ldots,29,\underbrace{30,31}_{{\cal S}}}^{\text{cyclic}},\overbrace{32,\ldots,114,\underbrace{115,\ldots,128}_{{\cal S}},129,\ldots,145}^{\text{noncyclic}} \,.
\end{align}
We address the problem of the numerical evaluation in section~\ref{sec:Implementation}.

\subsubsection{Weight-four pentagon functions}
\label{sec:explicit-repr:w4}

The 675 weight-four pentagon functions are given in terms of iterated integrals by
\begin{equation} \label{eq:4foldint}
\begin{aligned} 
f^{(4)}_i (X) = \, & \sum_{j_1,j_2,j_3,j_4} \tau^{(0)}_{i,j_1,j_2,j_3,j_4} [W_{j_1}, W_{j_2}, W_{j_3} , W_{j_4}]_{X_0}(X) \\ 
& + \sum_{j_1,j_2,j_3} \tau^{(1)}_{i,j_1,j_2,j_3}  [W_{j_1},W_{j_2},W_{j_3}]_{X_0}(X) \\ 
& + \sum_{j_1,j_2} \tau^{(2)}_{i,j_1,j_2}  [W_{j_1},W_{j_2}]_{X_0}(X) + \sum_{j_1} \tau^{(3)}_{i,j_1} [W_{j_1}]_{X_0}(X) +  \tau^{(4)}_i \,, 
\end{aligned}
\end{equation}
where $\tau^{(w)}_I$ for any set of indices $I$ denotes a weight-$w$ constant. The first 113 functions are cyclic, namely only the subset $\{ f^{(4)}_{i} \}_{i=1}^{113}$ is needed to express the cyclic permutations $D_4$~\eqref{eq:D4} of the planar master topologies. 
Starting with transcendental weight four the letters in $\mathcal{Z}$~\p{eq:Zlett} and $W_{198} = \sqrt{\Delta_5}$ begin to play a role in our choice of the pentagon functions. Among the cyclic pentagon functions, the letters $\mathcal{Z}$ are present only in the subset $\{ f^{(4)}_{i} \}_{i=107}^{112}$, while the letter $W_{198} = \sqrt{\Delta_5}$ is present in $f^{(4)}_{113}$ alone. Due to this choice, we expect the finite remainders of color ordered amplitudes to require only $\{ f^{(4)}_{i} \}_{i=1}^{106}$ out of the 675 weight-four pentagon functions,
as is the case for the results already available in the literature~\cite{Badger:2021nhg,Badger:2021ega,Abreu:2021asb}.
Beyond the cyclic sector, the letter $W_{198} =\sqrt{\Delta_5}$ is also present in the pentagon functions $\{ f^{(4)}_i\}_{i=665}^{675}$. We therefore expect that these 11 weight-four functions, together with the cyclic function $f^{(4)}_{113}$, are not required in the finite remainders.
Of course, since we want to be able to evaluate all the master integrals and not only the finite remainders, we consider all the 675 weight-four pentagon functions in what follows.

We now proceed to constructing a one-fold integral representation for the weight-four pentagon functions. Instead of expanding the weight-four pentagon functions in terms of iterated integrals as in~\cref{eq:4foldint}, we represent them as one-fold $\dd \log$ integrals of the weight-three pentagon functions,
\begin{align} 
f^{(4)}_i(X) = & \sum_{j,k} c_{i,j,k} \int^1_0 \dd t \,  \partial_t \log(W_j(t)) \,f^{(3)}_k(t) + \tau^{(4)}_i \,, \label{eq:f4=intf3}
\end{align}
where the weight-three pentagon functions $f^{(3)}_k$ and the alphabet letters are pulled back to $[0,1]$ by the integration contour $\gamma \in \mathcal{P}_{45}$,
\begin{align}
f_k^{(3)}(t) := f_k^{(3)}\left(X=\gamma(t)\right) \,, \qquad \qquad W_j(t) := W_j\left(X=\gamma(t)\right) \,.
\end{align}
By substituting the weight-three pentagon functions $f^{(3)}_k$ with their one-fold integral representations~\p{eq:1foldw3} in~\cref{eq:f4=intf3} we rewrite the weight-four functions as two-dimensional integrals over a triangular domain, $\{ (t,s): 0 \leq t \leq 1, 0\leq s \leq t \}$,
\begin{multline}
f^{(4)}_i(X) = \\
\sum_j \int^1_0 \dd  t \, \partial_t \log(W_j(t)) \left[
\sum_{k,l} c_{i,j,k,l} \int^t_0 ds\, \partial_s \log (W_k(s)) \, h^{(2)}_{l}(s)  + \tau^{(3)}_{i,j} \right] + \tau^{(4)}_i \,. \label{eq:2foldw4}
\end{multline} 
In order to obtain a one-fold integral representation, we swap the order of the integrations in~\cref{eq:2foldw4}~\cite{Caron-Huot:2014lda}. This way one of the integrations becomes trivial,
\begin{align}
L_j(s) := \int^1_s \dd t\, \partial_t \log(W_j(t)) \,, \label{eq:L}
\end{align}
and we obtain the desired expression in terms of one-fold integrals,
\begin{align}
f^{(4)}_i(X) = \sum_{j,k,l}  c_{i,j,k,l} \int^1_0 \dd  s\,\partial_s \log( W_k(s) ) \, L_j(s) \, h^{(2)}_{l}(s) + \sum_{j} \tau^{(3)}_{i,j} L_j(0) + \tau^{(4)}_i \,.
\label{eq:1foldw4}
\end{align} 
We rely on this representation of the weight-four pentagon functions for their numerical evaluation. In this view, we need to verify that the integrations in~\cref{eq:1foldw4} are well-defined. As argued for the analogous expression of the weight-three pentagon functions given by~\cref{eq:1foldw3}, the only possible pole singularities originate from the algebraic factors $\partial_s \log( W_k(s) )$. The spurious letters, $W_k \in {\cal S}$, may in fact vanish on the integration path $\gamma$. The weight-four pentagon functions however inherit the pole cancellation mechanism from the weight-three ones, see~\cref{eq:hvanish1,eq:hvanish2}. There is an additional complication with respect to the weight-three case.
Despite the pole singularity suppression, there are integrable logarithmic singularities, which arise from those terms in~\cref{eq:1foldw4} with $W_j \in {\cal S}$, so that $L_j(s)$ provides a logarithmic singularity.\footnote{In principle the factors of $\partial_s \log(W_k(s))$ and $L_j(s)$ in~\cref{eq:1foldw4} may contain distinct spurious letters, i.e.\ $W_k,\,W_j \in {\cal S}$ with $k\neq j$. However, we encounter such terms only with $j=k$.} We discuss the treatment of these integrable singularities in view of the numerical integration in section~\ref{sec:Implementation}.

The last piece of the weight-four pentagon functions which needs to be discussed are the logarithmic factors $L_j$ defined by~\cref{eq:L}. In appendix~\ref{app:alphpositivity} we study the behavior of the alphabet letters in the physical region ${\cal P}_{45}^+$. If the letter $W_j$ is real-valued and has constant sign along the path $\gamma$, then 
\begin{align}
L_j(s) = \log\left( \frac{W_j(X)}{W_j(\gamma(s))} \right)  \,.
\end{align}
The complex-valued letters, $\{W_{j}\}_{j=130}^{137} \cup \{W_{j}\}_{j=186}^{188}$, also do not pose any problem. They take values on the unit circle for any $X \in {\cal P}_{45}^+$ (see appendix~\ref{app:c-val-letters}). We can define their phases $\varphi_j(X)$, 
\begin{align}
\log(W_j(X)) = \ii  \varphi_j(X) \,,
\end{align}
to be continuous functions ranging in the interval $(0, 2\pi)$, 
\begin{align}
0<\varphi_j(X) < 2\pi \,,
\end{align}
so that the integration in~\cref{eq:L} is also trivial,
\begin{align}
L_j(s) = \ii  \varphi_j(X) - \ii  \varphi_j(\gamma(s))\,.
\end{align}
Finally, we argue that for the spurious letters $W_j \in {\cal S}$, which are all real-valued, we can choose
\begin{align} \label{eq:Ljspurious}
L_j(s) = \log \left| \frac{W_j(X)}{W_j(\gamma(s))} \right|.
\end{align}
Indeed, if a spurious letter $W_j(t)$ takes opposite signs at $t=1$ and $t=s$, $L_j(s)$ gets a branch contribution which depends on the prescription for the integration contour deformation. However, the branch contribution cancels out among the first and second term on the right hand side of~\cref{eq:1foldw4}. The weight-three pentagon function $f^{(3)}_k(t)$ accompanying $\partial_t \log(W_j(t))$ in~\cref{eq:f4=intf3} with spurious $W_j \in {\cal S}$ vanishes at $t=t_0$ for each $t_0$ such that $W_j(t_0)=0$, i.e.\     
\begin{align} \label{eq:zerofunw3}
\tau^{(3)}_{i,j}  + \sum_{k,l} c_{i,j,k,l} \int^{t_0}_0 ds \, \partial_s \log (W_{k}(s))\, h^{(2)}_{l}(s) = 0 \,, 
\end{align} 
thus suppressing the pole singularity originating from $\partial_t \log(W_j(t))$. This ensures the cancellation of the branch cuts of $L_j(s)$ in~\cref{eq:1foldw4}, which can then be completely ignored.
Note that the factors of $L_j(0)$ in~\cref{eq:1foldw4} corresponding to the linear spurious letters $W_j \in {\cal S}_{\ell}$, which vanish at the base point $X_0$, are absent in the weight-four pentagon functions. The remaining letters do not vanish at the base point, so that the corresponding $L_j(0)$'s are well-defined. 

So far we tacitly assumed that the weight-four pentagon functions $f^{(4)}_i(X)$ are evaluated at a phase-space point $X = \gamma(1)$ which is not in a zero locus of the spurious letters. This special case however deserves some care, as $L_j(s)$ is divergent if $W_j(X) = 0$. In fact, this singularity cancels out from the pentagon functions, which are not singular on any of the spurious surfaces $\{ X : W_j(X) = 0, \, W_j \in {\cal S} \}$. In order to show this, let us assume that the spurious letter $W_j$ vanishes on the path $\gamma(t)$ at $t = 1+\varepsilon$ with $|\varepsilon| \ll 1$, so that $|W_j(1)| \ll 1$. In other words, we assume that the integration along $\gamma$ ends in the proximity of a spurious singularity. Using~\cref{eq:zerofunw3} with $t_0 = 1+\varepsilon$ and~\cref{eq:Ljspurious}, we can rewrite the contributions to the right-hand side of~\cref{eq:1foldw4} with $j: \,  W_j \in{\cal S}$ as
\begin{equation} \label{eq:spQ-end-div}
\begin{aligned}
\sum_{j: W_j\in{\cal S}} \biggl\{ &  - \tau^{(3)}_{i,j} \log|W_j(X_0)|  - \sum_{k,l} c_{i,j,k,l} \int\limits^1_0 \dd  s\,\partial_s \log( W_k(s) ) \, \log|W_j(s)|\, h^{(2)}_{l}(s) \\
& -  \log|W_j(X)| \sum_{k,l}  c_{i,j,k,l} \int\limits^{1+\varepsilon}_1 \dd  s\,\partial_s \log( W_k(s) ) \, h^{(2)}_{l}(s) \biggr\} \,,
\end{aligned}
\end{equation}
where each term is manifestly finite as $\varepsilon \to 0$. Indeed, the second term contains an integrable logarithmic singularity, and the last term vanishes at $\varepsilon \to 0$. The remaining contributions to the right-hand side of~\cref{eq:1foldw4} with $j: \,  W_j \notin {\cal S}$ are clearly finite as well. As a result, the weight-four pentagon functions $f^{(4)}_i(X)$ are finite also when evaluated at a phase-space point $X$ where any of the spurious letters vanishes.

From the considerations above it follows that all the weight-four pentagon functions are well-defined in the entire physical scattering region ${\cal P}_{45}^+$. Some of them have bounded and explicitly finite integrands. Others, which involve $\dd \log$'s of spurious letters ${\cal S}$ in their one-fold integral representation~\p{eq:1foldw4}, demand additional work in view of the numerical integration. For this reason we isolate the latter in a minimal subset, as done already at transcendental weight three. This separation, on top of splitting the functions into cyclic and non-cyclic sectors, and isolating the letters $\mathcal{Z}$~\eqref{eq:Zlett} and $W_{198}=\sqrt{\Delta_5}$, gives the following classification of the weight-four pentagon functions $\{f^{(4)}_i\}_{i=1}^{675}$,
\begin{equation*}
\resizebox{0.98\hsize}{!}{$
\overbrace{1,\ldots,67,\underbrace{68,\ldots,106}_{{\cal S}}, \underbrace{107,\ldots,112}_{\mathcal{Z}},\underbrace{113}_{{\cal S}, \sqrt{\Delta_5}}}^{\text{cyclic}},\overbrace{114,\ldots,441,\underbrace{442,\ldots,664}_{{\cal S}}, \underbrace{665,\ldots,672}_{\sqrt{\Delta_5}}, \underbrace{673,\ldots,675}_{{\cal S},\sqrt{\Delta_5}} }^{\text{non-cyclic}}
$} \,.
\end{equation*}
\\

Now that we have expressions for all the one-mass pentagon functions that are well-defined throughout the physical region ${\cal P}_{45}^+$ we can move on to discussing their implementation in a \texttt{C++} library which allows for their fast and stable numerical evaluation. This is the topic of the next section.

\section{Numerical evaluation}
\label{sec:Implementation}

We implement the numerical evaluation of the one-mass pentagon functions in the \texttt{C++} library \texttt{PentagonFunctions++} \cite{PentagonFunctions:cpp},
which already provides the numerical evaluation of the massless pentagon functions from ref.~\cite{Chicherin:2020oor}.

The main idea of our approach follows the one from ref.~\cite{Chicherin:2020oor}. 
We implement the evaluation of the weight-one and two functions through their explicit representations given in \cref{sec:explicit-repr:w1,sec:explicit-repr:w2}.
For the weight-three and four functions we employ the one-fold integral representations derived in \cref{sec:explicit-repr:w3,sec:explicit-repr:w4}.
The one-fold integrals are computed numerically with the \emph{tanh-sinh} quadrature~\cite{Mori:1973}, which guarantees fast convergence
for the integrands that are analytic and bounded everywhere inside the integration domain, with the exception of the endpoints where integrable divergences can occur~\cite{Mori:1973,Bailey:2005}.
In \texttt{PentagonFunctions++} we use an adapted implementation of the tanh-sinh quadrature from \texttt{Boost C++}~\cite{BoostTanhSinh},
and we generate optimized code for the integrands with \texttt{FORM} \cite{Kuipers:2013pba,Ruijl:2017dtg}. 
We implement numerical evaluation in three fixed-precision floating-point types:  \emph{double},  \emph{quadruple}, and \emph{octuple}, approximately corresponding to
significands of 16, 32, and 64 decimal digits respectively. The last two floating-point types are available in \texttt{qdlib}~\cite{QDlib}.
We refer for more details to ref.~\cite{Chicherin:2020oor}.

In this section, we discuss the practical implications for our implementation of the two features of the one-mass five-particle phase space which are new with respect to the fully massless case.
The first new feature is the fact that the phase space $\mathcal{P}_{45}$, as an open subset of the affine space with coordinates~\eqref{eq:Mand}, is not starshaped, i.e.\
there is no phase-space point from which all other points are reachable by straight lines fully within $\mathcal{P}_{45}$ (see \cref{app:Shape} for details).
The second new feature is the existence of spurious singularities associated with the quadratic letters $\mathcal{S}_q$ that can occur anywhere along the integration path, as opposed to only at the initial point.

\subsection{Integration path}

As we show in \cref{app:Shape}, some points $X\in \mathcal{P}_{45}^+$ cannot be reached from $X_0$ by a line segment $\gamma \in \mathcal{P}_{45}^+$.
Therefore, instead of line segments, we consider integration paths $\gamma$ which are polygonal chains, i.e.\ connected series of line segments. Given a set of points $\{A_1,A_2,\ldots,A_n\}$ in the affine space, we denote by $[A_1,A_2,\ldots,A_n]$ the polygonal chain with vertices $A_1$,  $A_2$, ..., $A_n$.
To avoid explicit analytic continuation and crossing of physical singularities we require that $\gamma$ lies entirely within the physical phase space $\mathcal{P}_{45}^+$.

We begin by noticing that a line segment $\gamma_j(t) : [0,1] \to [X_0, X]$ lies within $\mathcal{P}_{45}^+$ if and only if it does not intersect the variety $\Delta_5=0$, namely if the pull back $\Delta_5\vert_{\gamma_j(t)}\neq 0$ for $t\in[0,1]$.
Therefore, using Sturm's theorem to count the number of roots of the quartic equation $\Delta_5\vert_{\gamma_j(t)}=0$, we can determine if any given line is inside $\mathcal{P}_{45}^+$ or not.
We can then construct a two-chain $\gamma = [X_0,X^\prime,X]$ with the following algorithm.
\begin{algorithm}[H]
\begin{enumerate}
  \item Generate a random point $X^\prime \in \mathcal{P}_{45}^+$ by the \texttt{RAMBO} algorithm~\cite{Kleiss:1985gy}.

  \item Reject the point and go to 1 if any of the following is true:
    \begin{enumerate}
        \item any of the spurious linear letters $\mathcal{S}_{\ell}$ change their sign on $[X^\prime,X]$;
        \item $X^\prime$ is close to the region boundary or to subvarieties of vanishing $\mathcal{S}_{\ell}$; \footnotemark    
    \end{enumerate}

  \item If both line segments $[X_0, X^\prime]$ and $[X^\prime, X]$ lie within $\mathcal{P}_{45}^+$, accept $X^\prime$ and terminate, else go to 1.
\end{enumerate}
\caption{Monte Carlo algorithm for the construction of a polygonal chain within $\mathcal{P}_{45}^+$.}
\label{alg:algpath}
\end{algorithm}
\footnotetext{The closeness is determined by a technical cutoff on the absolute values of the relevant letters.}
Here step 2 is designed to avoid potential numerical instabilities arising from the path unnecessarily entering near-singular configurations, as well as to avoid unnecessary crossing of spurious singularities.
Some care should be taken when the coefficients of the $\Delta_5(t)$ polynomial become small and the number of roots cannot be reliably calculated from Sturm's theorem in step 3.
In these cases we reject the point and try another one.

This simple strategy allows us to quickly check thousands of points and find a suitable polygonal two-chain. 
In principle, more line segments may be required and the presented algorithm can be straightforwardly extended.
However, in practice we observe that two connected line segments are sufficient provided the base point $X_0$ is given by \p{eq:basepoint}, and rarely more than 100 attempts are needed to find a suitable point $X^\prime$.
We thus conjecture that any point of $\mathcal{P}_{45}^+$ can be reached from $X_0$ by at most two connected line segments.
If \texttt{PentagonFunctions++} ever encounters a point for which this is not the case, an error message is generated and the evaluation is aborted. We observe that, from a sample of $\order{10^6}$ randomly generated points, only a small fraction (few percents) is not visible from $X_0$. We can view this as a statistical measure of non-starshapedness of the physical region.

\subsection{Spurious singularities}

In \cref{sec:explicit-repr:w3,sec:explicit-repr:w4} we demonstrated that all integrands of the one-fold integrals in \cref{eq:1foldw3,eq:1foldw4} 
are either finite or have integrable divergences on subvarieties where any of the spurious letters $W_i\in\mathcal{S}$ is vanishing.
Hence we showed that the integrals involving spurious divergences are well-defined from the mathematical point of view.
Nevertheless, their numerical evaluation does require special treatment to ensure adequate numerical stability.

Whenever one of the quadratic spurious letters $W_i \in {\cal S}_q$ has opposite signs at the base point $X_0$ and at the evaluation point $X$,
the integration path $\gamma(t)$ has to cross the spurious singularity $W_i = 0$ for some $t_0\in[0,1]$.
The corresponding one-form $\dd{\log W_i}$ then has a simple pole at $t_0$, which is compensated by the corresponding multiplier $h(X)$ which vanishes at $t=t_0$ (see \cref{sec:explicit-repr:w3}).
This commonly causes significant loss of precision due to numerical cancellations.
We address this problem by evaluating the contribution proportional to $\dd{\log W_i}$ in the neighborhood where $W_i$ is small as 
\begin{equation} \label{eq:spur-series-repr}
  \qty(h(X) \dd{\log W_i}) \vert_{\gamma_j(t)} = \left( h_{1}(t) + h_{2}(t) \,W_i(t) + \order{W_i^2} \right) \dd{W_i \vert_{\gamma_j(t)}},
\end{equation}
where $h_1,~ h_2$ are the first non-vanishing coefficients of the series expansion of $h(X)$ in $W_i$.
Therefore the spurious poles are always canceled analytically.
We choose the threshold for when to switch into the series expansion such that the $ \order{W_i^2} $ corrections are estimated to be smaller than the roundoff error at given precision.

At weight three the coefficients $h_1,~ h_2$ are explicitly finite by construction.
At weight four they can contain logarithmic singularities of the form $\log\abs{W_i}$ originating from \cref{eq:Ljspurious}.
These singularities are integrable.
Nevertheless, they violate the assumptions of the tanh-sinh quadrature if they are encountered anywhere except at the endpoints.
If left unattended, they spoil the expected double-exponential convergence rate and the associated error estimates.
To avoid this issue, we further subdivide the integration chain at the locations of the spurious singularities such that they stand at the endpoints by construction. We do this on-demand for each pentagon function individually.
As in the case of linear spurious singularities at $X_0$,
we must additionally ensure that, close to the spurious singularity, the vanishing letters are evaluated in a numerically stable manner,
and that the rounding errors do not cause the integrand to be evaluated exactly at the singularity.
We accomplish this by evaluating the letters that can vanish along a line segment $\gamma_j$ as
\begin{equation} \label{eq:Wspur-on-path-eval}
  W_i\vert_{\gamma_j(t)} = c^{(1)}_{i;\gamma_j} ~ (t-t_0) + c^{(2)}_{i;\gamma_j}  ~ (t-t_0)^2 \,, 
\end{equation}
where $t_0$ is a solution of $W_i\vert_{\gamma_j(t)}=0$, and the coefficients $c^{(k)}_{i;\gamma_j}$ are computed explicitly from the letter's derivatives.
If a letter $W_i \in \mathcal{S}_q$ vanishes more than once along the same line segment, we construct multiple representations of the form \eqref{eq:Wspur-on-path-eval} for each of the singularity locations $t_k$.
Then for any $t \in \gamma_j$ we select the representation which corresponds to the singularity that is the closest, i.e.\ such that $\abs{t-t_k}$ is minimized.

The same considerations apply also to the case of linear spurious singularities $W_i \in \mathcal{S}_{\ell}$ except that their location is always fixed to $t_0=0$ by construction.
The representation in \cref{eq:spur-series-repr} remains unchanged, while in \cref{eq:Wspur-on-path-eval} the coefficients $c^{(2)}_{i;\gamma_j}=0$.

Finally, let us note that the same considerations apply to the case when the point $X$ itself is close to the zero locus $W_i=0$.
As far as floating-point evaluations are concerned, no extra precautions are needed in this case, as long as $W_i(X)\neq 0$ \emph{exactly}.
Indeed, the remaining spurious divergence is only logarithmic, hence it does not become large even for $W_i$ as small as machine epsilon.
If a specifically fine tuned phase-space point is of interest, we refer to the representation in \cref{eq:spQ-end-div}.

\subsection{Performance}

To characterize the performance of our implementation we sample all one mass pentagon functions on points from a physical phase-space distribution that is representative of the expected phenomenological applications.
For the purpose of demonstration, we construct a phase-space distribution by requiring optimal Monte Carlo integration of the leading order $pp\to e \bar{\nu}_e jj$ production cross section at $\sqrt{s} = 13 \text{TeV}$ with \texttt{Sherpa 2.2} \cite{Bothmann:2019yzt}.
We define the phase space by requiring two anti-$k_\text{T}$ jets with $R=0.4$ \cite{Cacciari:2011ma}, and we apply rather loose cuts,
\begin{equation}
  p_{\text{T}}^{l} > 20 \text{GeV}\,, \quad 
  p_{\text{T}}^{j} > 25 \text{GeV}\,, \quad 
  M^{W} > 20 \text{GeV}\,, \qquad 
  \abs{\eta^{e}} < 2.5\,, \quad 
  \abs{\eta^{j}} < 3\,, \quad 
\end{equation}
where $p_{\text{T}}^{l}$ and $p_{\text{T}}^{j}$ are the transverse momenta of the leptons and jets respectively, $M^{W}$ is the dilepton invariant mass, and
$\eta^{e},\eta^{j}$ are the pseudo-rapidities of the electron and jets respectively.
The renormalization scale is chosen to be half the scalar sum of the transverse momenta of all final-state particles. 

\begin{figure}[ht]
  \centering
  \includegraphics[width=0.8\textwidth]{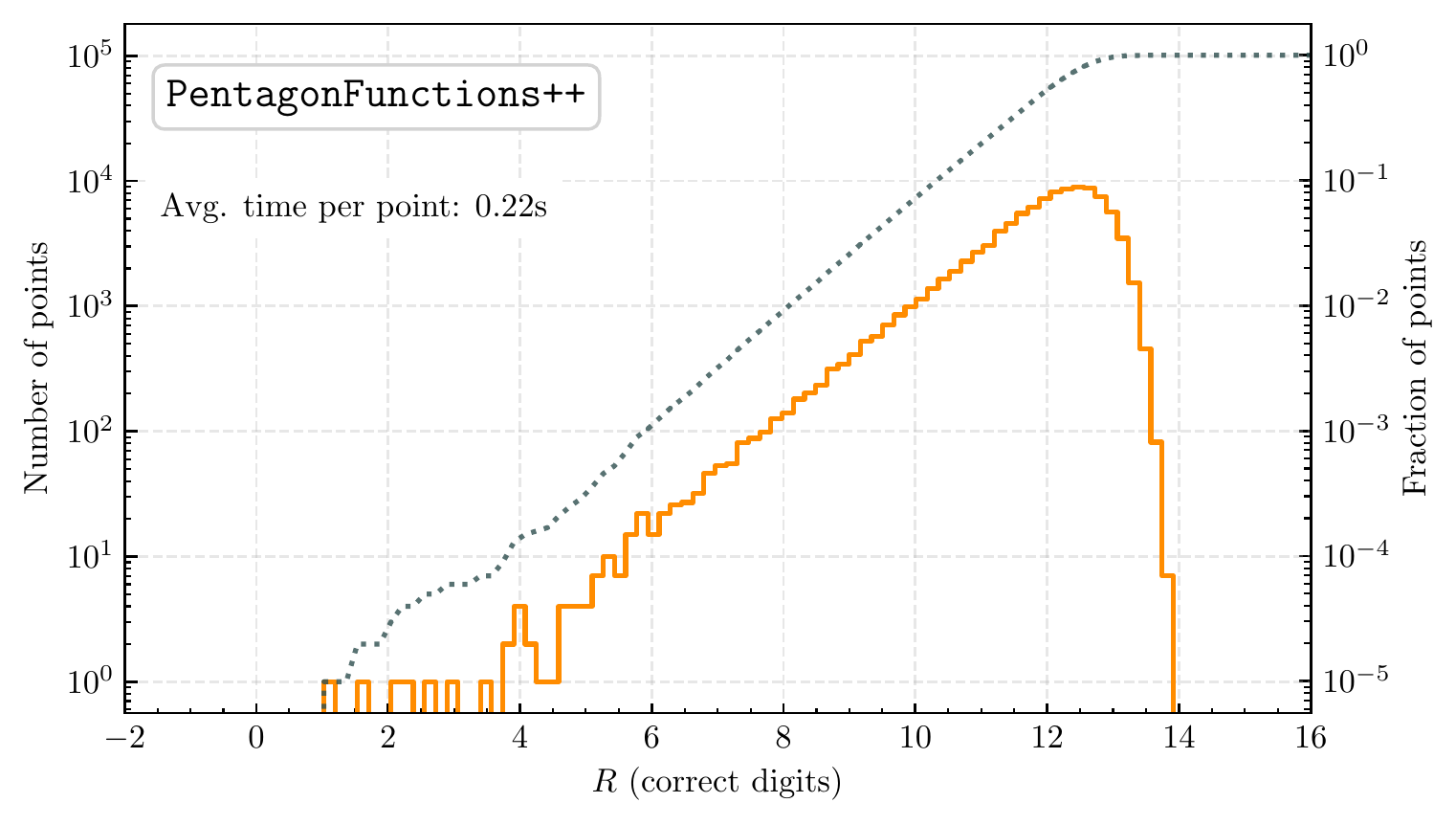}
  \caption{
    Distribution of logarithmic relative error of the one-mass pentagon functions (see \cref{eq:mindigits}) sampled on $10^5$ kinematical points from a typical physical phase space (solid orange line).
    The cumulative distribution is displayed by the dashed line.
    The average evaluation time of all one-mass pentagon functions in double precision on a single thread is estimated in a run where kinematical points are evaluated in parallel on a server with \texttt{Intel(R) Xeon(R) Silver 4216 CPU @ 2.10GHz}.
  }
  \label{fig:num_stab}
\end{figure}

We evaluate all pentagon functions in double and quadruple precision on $10^5$ phase-space points from this distribution.
We characterize the accuracy of the double-precision evaluation $\hat{f}^{(w)}_{i;\text{double}}(X)$ of a pentagon function on a kinematical point $X$ by the logarithmic relative error $r_{w,i}$,
\begin{equation}\label{eq:digits}
  r_{w,i}(X) = -\log_{10} \left| \frac{\hat{f}^{(w)}_{i;\text{double}}(X)}{\hat{f}^{(w)}_{i;\text{quad}}(X)} - 1 \right|,
\end{equation}
where $\hat{f}^{(w)}_{i;\text{quad}}(X)$ is the numerical evaluation of the same function in quadruple precision.
For convenience we will refer to this quantity as \emph{correct digits}.
We define the smallest number of correct digits (i.e.\ worst precision) among all pentagon functions at the kinematical point $X$ as
\begin{equation} \label{eq:mindigits}
  R(X) = \min_{w,i} \left[r_{w,i}(X)\right], \qquad w,i \in \{\text{all functions up to weight four}\} .
\end{equation}
We display the distribution of $R(X)$ over the phase space as well as the average evaluation time in \cref{fig:num_stab}.
We observe excellent precision in the bulk of the phase space.
A few of the phase-space points in the tail with low number of correct digits can be associated to the fact that the evaluation of $\trfive$ on those points becomes ill-conditioned.
Comparing to the case of the massless pentagon functions in \cite{Chicherin:2020oor},
the evaluation time is even lower despite the more complex phase space and the larger number of functions.
We attribute this to the implementation details of the \texttt{C++} library.
This fact does however demonstrate that the overhead from integrating over multiple line segments is indeed negligible.

To characterize the scaling of our method with precision, we show in \cref{tab:timings} the timings of evaluation of all one-mass pentagon functions in double, quadruple, and octuple precision on a typical phase-space point. 
The number of correct digits is calculated by comparing with the results obtained from \texttt{DiffExp} as described in \cref{sec:InitialValues}.

\begin{table}[h]
  \centering
  \begin{tabular}{ccc}
    \toprule
    Precision & Correct digits & Timing (s) \\
    \midrule
    double & 12  & 0.19 \\
    quadruple & 28 & 159  \\
    octuple & 60 & 1695  \\ 
    \bottomrule
  \end{tabular}
  \caption{Evaluation times of all one-mass pentagon functions on a typical phase-space point. The evaluations are performed on a single thread of \texttt{Intel(R) Xeon(R) Silver 4216 CPU @ 2.10GHz}.}
  \label{tab:timings}
\end{table}

The termination condition for the numerical one-fold integration for the benchmarks in this section is chosen as specified by the default settings of \texttt{PentagonFunctions++}: it is picked in such a way that the precision of the integration result is only limited by rounding errors in the integrand and the condition number 
which characterizes the sensitivity of the evaluation result to small perturbations in the input.
One can therefore adjust this threshold to improve average evaluation time if lower precision in the bulk is sufficient for the desired applications.
For example, it might be useful if quadruple precision is necessary only to overcome rounding errors in the integrand, while the target precision is still that of double precision.

Finally, let us emphasize that in this section we considered the complete set of one-mass pentagon functions sufficient to express any two-loop five-point scattering amplitude with four massless and one massive external momenta.
It is however expected that not all of the functions contribute in particular applications.
For instance, it has been observed that the functions involving certain subsets of letters drop out from suitably defined finite quantities \cite{Chicherin:2020umh,Badger:2021nhg,Badger:2021ega}.
Moreover, some scattering processes require only the subset of pentagon functions corresponding to the cyclic permutations (see e.g.\ \cite{Badger:2021nhg}). Both of these properties are manifest in our function basis.
Hence, the results presented in this section can be considered as the worst-case scenario analysis.

\subsection{Validation}

We validated our results against three sources: the function basis of ref.~\cite{Badger:2021nhg}, the benchmark values for the pure master integrals in ref.~\cite{Abreu:2020jxa}, and the numerical evaluation of the master integrals using \texttt{DiffExp}. 

We verified that the basis functions of ref.~\cite{Badger:2021nhg}, which cover only the cyclic permutations $D_4$ of the planar integral families, can be expressed as (graded) polynomials in our cyclic one-mass pentagon functions, meaning that there is full compatibility between the two bases. We made use of this relation to cross-check the numerical values of the cyclic one-mass pentagon functions obtained with \texttt{PentagonFunctions++} against those of the functions of ref.~\cite{Badger:2021nhg} in several kinematic points, including some in $2\to3$ physical scattering regions different from $\mathcal{P}_{45}$. We evaluated the one-mass pentagon functions at the latter points in double precision as discussed in~\cref{sec:Permutations}, using the analytic permutations of the pentagon functions provided in the ancillary files~\cite{supplemental}. We found full agreement.

Reference~\cite{Abreu:2020jxa} provides the numerical values of the pure master integrals of the planar families in the standard orientation of the external legs in six kinematic points, one for each distinct $2\to3$ physical scattering region with the massive particle in the final state. We reproduced these values through the one-mass pentagon functions evaluated with octuple precision and crossed to the scattering regions different from $\mathcal{P}_{45}$ as discussed in~\cref{sec:Permutations}.

Finally, we used \texttt{DiffExp} to evaluate numerically all master integrals at several kinematic points in $\mathcal{P}_{45}$, including some close to spurious singularities. We used the benchmark values of ref.~\cite{Abreu:2020jxa} as initial values. We evaluated the pure master integrals of the permuted families by permuting the evaluation point as in~\cref{eq:gsigmaX} and flipping the sign of the parity-odd integrals for the odd-signature permutations. We then matched the expressions of the pure master integrals in terms of the one-mass pentagon functions with their numerical values, and solved the ensuing system of equations to get the values of the pentagon functions. The fact that this system of equations admits a unique solution confirms the correctness of the expressions of the integrals in terms of pentagon functions. We then cross-checked the resulting values against the evaluations done with \texttt{PentagonFunctions++} in octuple precision, finding full agreement to the expected accuracy.

\subsection{Usage}

The library \texttt{PentagonFunctions++} can be used either as a \texttt{C++} library or through its \texttt{Mathematica} interface implemented by 
the package \verb|PentagonFunctions`|. 
We give here only a brief demonstration and refer to the documentation supplied together with the library distribution \cite{PentagonFunctions:cpp}.

The library accepts as input the kinematical point given by the six Mandelstam invariants in \cref{eq:Mand}.
The latter are assumed to be in ratios to the regularization scale, see \cref{eq:integral-famililies}.
The pentagon functions are evaluated in their analyticity region $\mathcal{P}_{45}^{+}$ defined in \cref{eq:P45pm}, which means that $\Im \trfive>0$ by construction.
Function values for parity-conjugated points with $\Im \trfive <0$ can be obtained by manually changing the sign of the parity-odd functions, which we list explicitly in the supplementary file \texttt{pfuncs\_charges.m}~\cite{supplemental}.

The interface of the library is designed in such a way that for each needed pentagon function the user must first obtain a callable function object or \emph{evaluator}. 
In the \texttt{C++} interface the evaluator of numerical type \verb|T| is obtained by creating an instance of \verb|FunID<KinType::m1>| type, and invoking 
its template method \verb|get_evaluator<T>()|. For example, with the code
\begin{lstlisting}[language=c++,numbers=left]
    FunID<KinType::m1> fID(3,14);
    auto fobj = fID.get_evaluator<double>();

    auto result = fobj(X);
\end{lstlisting}
one obtains a double-precision evaluator \verb|fobj| for the function $f^{(3)}_{14}$.
In line 4 the function is evaluated on a point $X$ specified as in \cref{eq:Mand}.
It is worth noting that the created evaluators are completely independent of each other and will each evaluate only the requested pentagon functions. 
This means, for instance, that the user can choose to evaluate only a needed subset of pentagon functions.

Similarly, in the \texttt{Mathematica} interface this can be achieved with
\begin{lstlisting}[language=Mathematica,numbers=left]
  evaluator = StartEvaluatorProcess["m1", {F[3,14]}, "Precision" -> "double"];

  result = EvaluateFunctions[evaluator, X];
\end{lstlisting}
Here the second argument is a list of required functions, and the last argument can be used to select the required precision.
In the \texttt{Mathematica} interface the functions are evaluated in parallel using all available threads. 
In line 3 the evaluation result is returned as replacement rules.

In both interfaces \verb|m1| is used as an identification for the set of one-mass pentagon functions introduced in this paper,\footnote{
  The massless pentagon functions from ref.\ \cite{Chicherin:2020oor} can be accessed by using \texttt{m0} instead.
}
and the evaluator objects can be used for evaluations on any number of subsequent phase-space points.
For more details we refer to the examples provided with the library distribution.

\section{Conclusions}
\label{sec:Conclusions}

We constructed a basis of transcendental functions which allows to express any planar one- and two-loop five-particle Feynman integral with a single external massive leg up to transcendental weight four. This is sufficient for the computation of the planar two-loop corrections to cross sections.
We call the functions in our basis one-mass pentagon functions. Our results are valid throughout the entire physical phase space for any $2\to 3$ scattering process with production of a massive particle. We achieved this by constructing expressions for the one-mass pentagon functions which are well-defined in a particular $2\to3$ channel, and by providing the expressions of the master integrals in terms of our function basis for all permutations of the external massless legs. 
We expressed the one-mass pentagon functions in terms of logarithms and dilogarithms up to transcendental weight two. For the weight three and four functions, instead, we constructed a one-fold integral representation which is well suited for the numerical integration.
We presented a public \texttt{C++} library~\cite{PentagonFunctions:cpp} which allows to evaluate numerically the one-mass pentagon functions in a stable and fast way. 
The supplemental material can be obtained from the \texttt{git} repository~\cite{supplemental}.

The results presented in this paper open the door for broad phenomenological applications, and constitute a crucial step forward towards the computation of NNLO QCD predictions for several processes which feature prominently in the LHC physics program, such as the production of a Higgs or an electroweak vector boson in association with two jets. Indeed, the one-mass pentagon functions have already been instrumental in the recent calculation of the planar two-loop amplitudes for four partons and an electroweak vector boson~\cite{Abreu:2021asb}, and we envisage that they will facilitate calculations of other processes with similar scattering kinematics in the near future.

More broadly, we expect that the approach we adopted in this paper will help to tackle many further problems in the future. In primis, we envisage that it will be possible to classify the non-planar one-mass pentagon functions in a completely analogous way, once the canonical DEs for all the non-planar integral families are available (see refs.~\cite{Papadopoulos:2019iam,Abreu:2021smk} for progress in this direction). Similarly, the expressions for the one-mass pentagon functions may be generalized to the $1 \to 4$ decay region, which would for instance enable a two-loop description of $e^+e^- \to 4j$ at lepton colliders. More generally, our approach can be useful in other multi-scale applications, in particular those where rationalizing all square roots is very difficult or even impossible.

\acknowledgments
We are grateful to Simon Badger and Heribertus Bayu Hartanto for insightful discussions, and to Robin Br\"user, Johannes M. Henn, Leila Maestri and Tiziano Peraro for useful suggestions at the initial stages of this work. We thank Samuel Abreu, Simon Badger and Johannes M. Henn for comments on the manuscript.
The diagrams in figures~\ref{fig:families} and~\ref{fig:1Lsquared} have been drawn with \textsc{JaxoDraw}~\cite{BINOSI200476}. This project has received funding from the European Union's Horizon 2020 research and innovation programmes \textit{Novel structures in scattering amplitudes} (grant agreement No 725110), and \textit{High precision multi-jet dynamics at the LHC} (grant agreement No 772009). DC is supported by the French National Research Agency in the framework of the \textit{Investissements d’avenir} program (ANR-15-IDEX-02).
DC and SZ gratefully acknowledge the computing resources provided by the Max Planck Institute for Physics.

\appendix

\section{Gram determinants and the physical region}
\label{app:GramDet}

In this section we discuss in detail the properties of the Gram determinants which are relevant to define the physical scattering region and to determine the signs of the letters of the alphabet. Our discussion essentially follows refs.~\cite{Byers:1964ryc,POON1970509,Byckling:1971vca}, specializing to the case of massless particles. We begin by defining the Gram determinant of two sets of momenta,
\begin{align} \label{eq:Gram1}
G(q_{i_1} ,\ldots, q_{i_n} ; q_{j_1} ,\ldots, q_{j_n} ) = \det\, \mathcal{G}(q_{i_1} ,\ldots, q_{i_n} ; q_{j_1} ,\ldots, q_{j_n} ) \,,
\end{align}
where $\mathcal{G}$ is the Gram matrix,
\begin{align}
\mathcal{G}(q_{i_1} ,\ldots, q_{i_n} ; q_{j_1} ,\ldots, q_{j_n} )  = \begin{pmatrix} q_{i_1} \cdot q_{j_1} & \cdots & q_{i_1} \cdot q_{j_n} \\ \vdots & \ddots & \vdots \\ q_{i_n}\cdot q_{j_1} & \cdots & q_{i_n}\cdot q_{j_n} \end{pmatrix} \,.
\end{align}
We call the number of momenta in the two sets, $n$ in~\cref{eq:Gram1}, the \textit{order} of the Gram determinant. 
The Gram determinant is invariant under the exchange of the two sets of momenta, which corresponds to a transposition of the Gram matrix.
We further introduce a short-hand notation for the Gram determinants of the external momenta $\{ p_i \}_{i=1}^5$,
\begin{align}
& G_{i_1 \ldots i_n}^{j_1 \ldots j_n} = G(p_{i_1} ,\ldots, p_{i_n} ; p_{j_1} ,\ldots, p_{j_n} ) \,, \\
& G_{i_1 \ldots i_n} = G_{i_1 \ldots i_n}^{i_1 \ldots i_n}  \,.
\end{align}
The Gram determinant vanishes if the momenta in any of the two sets are linearly dependent, and is invariant under shifts of any of the momenta by any linear combination of the other momenta in its argument. As a result, when considering four-dimensional external momenta, the Gram determinants with order $n \ge 5$ vanish.
Moreover, the order-$n$ Gram determinant is invariant under any permutation $\sigma \in {\cal S}_n$ of the momenta in its argument, 
\begin{align}
G_{i_1 \ldots i_n}^{j_1 \ldots j_n} = G_{i_{\sigma(1)} \ldots i_{\sigma(n)}}^{j_{\sigma(1)} \ldots j_{\sigma(n)}}\,.
\end{align} 
Finally, Sylvester's determinant identity implies a number of relations among Gram determinants with different orders which will be useful in the following discussion,
\begin{align}
\label{eq:SylvId1}
& G_i \, G_{ijk} = G_{ij} G_{ik} - \left(G^{ij}_{ik} \right)^2 \,, \\
\label{eq:SylvId2}
& G_{kl} G_{ijkl} = G_{ikl} G_{jkl} - \left(G_{ikl}^{jkl}\right)^2 
\,, \\ 
& \left(G_i\right)^2  G_{ijkl} = \det 
\begin{pmatrix}
  G_{ij} & G^{ij}_{ik} & G^{ij}_{il} \\
  G^{ij}_{ik} & G_{ik} & G^{ik}_{il} \\
  G^{ij}_{il} & G^{ik}_{il} & G_{il}
\end{pmatrix} \,. \label{eq:SylvId3}
\end{align}

Now that we have stated the main properties of the Gram determinants we can move on to discussing how they constrain the physical region for five-particle scattering~\cite{Byers:1964ryc,POON1970509,Byckling:1971vca}. In particular, we prove the constraints given by~\cref{eq:GramDetConstraints}. Let us consider the Gram determinants $G_{i_1 \ldots i_n}$, with $n=1,\ldots,4$ and distinct indices $i_1,\ldots, i_n$. 
The corresponding Gram matrix $\mathcal{G}(p_{i_1},\ldots,p_{i_n})$ is symmetric and thus has $n$ real eigenvalues. Clearly if any of the eigenvalues is equal to zero, the determinant $G_{i_1 \ldots i_n}$ also vanishes. We can prove that the Gram matrix $\mathcal{G}(p_{i_1},\ldots,p_{i_n})$ is either identically zero ---~which is of no interest~--- or it has exactly one positive and $n-1$ non-positive eigenvalues. Since the Gram determinant $G_{i_1,\ldots,i_n} $ is the product of all eigenvalues of the corresponding Gram matrix, this property straightforwardly implies the constraints given by~\cref{eq:GramDetConstraints}, which we rewrite here for the convenience of the readers,
\begin{align} \label{eq:GramDetConstraintsApp}
G(p_1) > 0 \,, \qquad G(p_i, p_j) < 0 \,, \qquad G(p_i,p_j,p_k) > 0 \,, \qquad G(p_i,p_j,p_k,p_l) < 0 \,.
\end{align}

We begin by proving that the Gram matrix $\mathcal{G}(p_{i_1},\ldots,p_{i_n})$ cannot have more than one positive eigenvalue. If it had two positive eigenvalues, the corresponding eigenvectors would be real linear combinations of $p_{i_1},\ldots,p_{i_n}$ with the following properties: linear independence, orthogonality, and time-likeness. This is impossible in the $(1,3)$ Minkowski space-time metric. Therefore either all $n$ eigenvalues of $\mathcal{G}(p_{i_1},\ldots,p_{i_n})$ are non-positive, or one eigenvalue is positive and $n-1$ are non-positive.

Secondly, we prove that the Gram matrix $\mathcal{G}(p_{i_1},\ldots,p_{i_n})$ has at least one positive eigenvalue or it vanishes identically.
This follows from physical constraints on the kinematics. In particular, we assume that the momentum $p_1$ is time-like, $p_1^2 > 0$. The sum of the eigenvalues of $\mathcal{G}(p_{i_1},\ldots,p_{i_n})$ equals its trace, i.e.\ $\sum_{k=1}^n p_{i_k}^2$. Then either $i_k = 1$ for some $k$, in which case the trace of the Gram matrix evaluates to $p_1^2$ and is thus positive, or no index is equal to $1$, in which case the trace vanishes. The sum of the eigenvalues is thus non-negative. As a result, either all the eigenvalues are zero or at least one eigenvalue is positive. 
Since the Gram matrix $\mathcal{G}(p_{i_1},\ldots,p_{i_n})$ can have at most one positive eigenvalue, this implies that it has exactly one positive and $n-1$ non-positive eigenvalues, or that it vanishes. q.e.d.

The study of the positivity of the alphabet letters in appendix~\ref{app:alphpositivity} makes use of another useful identity,
\begin{align} \label{eq:idG3}
G(p_1,p_i + \alpha p_j + \beta p_k) < 0\,, \quad \forall \alpha, \beta \in \mathbb{R} \,,
\end{align}
with distinct indices $i,j,k$ taken from $\{ 2,3,4,5\}$. This relation holds in the physical scattering regions, i.e.\ where the Gram-determinant constraints~\eqref{eq:GramDetConstraintsApp} are satisfied.
We prove this by showing that $G(p_1,p_i + \alpha p_j + \beta p_k)$ is a non-positive polynomial in $\alpha$ and $\beta$,
\begin{align}
G(p_1,p_i + \alpha p_j + \beta p_k) = \alpha^2 G_{1j} + 2 \alpha \beta G_{1k} + \beta^2 G_{1k} + 2 \alpha G^{1i}_{1j} + 2 \beta G^{1i}_{1k} +G_{1i} \,.
\end{align}
Using the identity~\eqref{eq:SylvId1} we see that the determinant of the Hessian matrix of this polynomial is non-negative,
\begin{align}
G_{1j} G_{1k} - \left(G^{1j}_{1k}\right)^2 = G_{1} G_{1jk} > 0 \,.
\end{align}
This, together with the physical constraint that $G_{1j}<0$, implies that the polynomial has a global maximum where the first derivatives in $\alpha$ and $\beta$ vanish. With the help of \p{eq:SylvId1} and \p{eq:SylvId3}, we rewrite the maximum in terms of the Gram determinants so that it is explicitly non-positive,
\begin{align}
\underset{\alpha,\beta}{\max} \, G(p_1,p_i + \alpha p_j + \beta p_k) = \frac{G_{1} \, G_{1ijk}}{G_{1jk}} < 0 \,,
\end{align}
which implies the inequality~\eqref{eq:idG3}. q.e.d.

In order to understand the implications of the Gram-determinant constraints~\eqref{eq:GramDetConstraintsApp} in terms of the scalar invariants it is useful to spell out the various Gram determinants. We begin with the order-one Gram determinants. Only one of them is non-vanishing, whereas the other four vanish identically,
\begin{align}
G_{1} = p_1^2 > 0 \,, \qquad \quad G_{j} = 0 \ \forall j=2,3,4,5 \,.
\end{align}
Higher order Gram determinants vanish only in degenerate configurations of the momenta corresponding to the boundaries of the physical region defined by~\cref{eq:pipjConstraints} (see also \cref{footnote:degenerate-momenta}). The order-two Gram determinants are given by
\begin{align}
G_{ij} = - (p_i \cdot p_j)^2 < 0 \qquad \forall i,j=1,\ldots,5\,, i\neq j\,.
\end{align}
Clearly the $G_{ij}$'s, for distinct $i$ and $j$, can vanish only where $p_i \cdot p_j = 0$, which indeed defines the boundary of the physical regions. 
The order-three Gram determinants have different expressions depending on whether one of the indices is equal to $1$,
\begin{align}
& G_{ijk} = \frac{1}{4} s_{ij} s_{ik} s_{jk} \,, \\
& G_{1ij} = \frac{s_{ij}}{4} \left((s_{1i}-p_1^2)(s_{1j}-p_1^2)-p_1^2 s_{ij}\right) \,, \label{eq:Gord3}
\end{align}
where $i,j,k$ take distinct values in $\{2,3,4,5\}$. In order to show that the order-three Gram determinants vanish only on the boundaries of the physical regions we make use of~\cref{eq:SylvId2}. Setting $G_{ikl} = 0$ in~\cref{eq:SylvId2} in fact implies that
\begin{align}
G_{kl} G_{ijkl} = - \left(G_{ikl}^{jkl}\right)^2 \,.
\end{align}
Because of the Gram-determinant constraints~\eqref{eq:GramDetConstraintsApp} in the physical scattering regions, the left-hand side is non-negative, while the right-hand side is non-positive. Hence, both sides of the equation vanish. The vanishing of the order-three Gram determinant $G_{ikl} = 0$ thus implies that either $G_{kl}=0$ or $G_{ijkl}=0$. Both cases correspond to boundaries of the physical regions. We have already proven this above for the order-two Gram determinants and we discuss it presently for the order-four ones. The latter are all proportional to $\Delta_5 = \trfive^2$,
 \begin{align}
 G_{ijkl} = \frac{\Delta_5}{16}\,,
 \end{align}
 which clearly can only vanish at the boundary $\Delta_5 = 0$ of the physical region (see also \cref{footnote:degenerate-momenta}).

Finally, let us establish the inequalities in~\cref{eq:P45}, which define the $s_{45}$ scattering region in terms of the Mandelstam invariants $s_{ij}$ and $p_1^2$. We begin by rewriting the physical constraints on the scalar products~\p{eq:pipjConstraints} in terms of the Mandelstam variables,
\begin{equation} \label{eq:pipjConstraintsToMandelst}
\begin{aligned} 
& s_{12} > p_1^2\,, s_{13} > p_1^2 \,, s_{23}>0 \,, s_{45}>0 \,, s_{24} < 0 \,,  \\ 
& s_{34} < 0 \,, s_{25} < 0 \,, s_{35} < 0 \,, s_{14} < p_1^2 \,, s_{15} < p_1^2 \,.
\end{aligned}
\end{equation} 
We see that the latter are less strict than the constraints given in~\cref{eq:P45}. However, in view of \p{eq:pipjConstraintsToMandelst},
\begin{align} 
s_{45} = s_{12} + s_{23} + (s_{13}-p_1^2) > p_1^2 \,. \label{s45p1sq}
\end{align} 
Then we consider the following order-three Gram determinant,
\begin{equation}
\begin{aligned}
0 < 4\,G_{145} 
&= -s_{45} \left[ s_{23} (p_1^2 - s_{15}) + s_{15}^2 + s_{15} (s_{45}-p_1^2)\right] \\
&= -s_{45}\left[ s_{23} (p_1^2 - s_{14}) + s_{14}^2 + s_{14} (s_{45}-p_1^2)\right] \,.
\end{aligned}
\end{equation}
Its positivity, together with the constraints~\p{eq:pipjConstraintsToMandelst} and~\p{s45p1sq}, implies that
\begin{align}
s_{14} <0 \,,\quad s_{15} < 0 \,.  \label{s14ands15} 
\end{align}
Finally, completing the constraints~\p{eq:pipjConstraintsToMandelst} with those given in~\cref{s45p1sq,s14ands15} gives the stricter inequalities~\p{eq:P45}.

\section{Shape of the physical phase space}
\label{app:Shape}

In this appendix we discuss the convexity properties of the physical phase space ${\cal P}_{45}$. The latter is an open subset of the affine space with coordinates $\left( p_1^2, s_{12},s_{23},s_{34},s_{45},s_{15} \right)$ carved out by the inequalities~\p{eq:P45}. 

Given two points $x, y$ in some affine space $\mathcal{A}$, we say that $y$ is \emph{visible} from $x$ (or vice versa) if the line segment connecting $x$ and $y$ lies in $\mathcal{A}$.
First of all, we note that one can easily find two points within ${\cal P}_{45}$ which are not visible from each other, i.e.\ $\mathcal{P}_{45}$ is not convex. 
A more refined question about the geometry of ${\cal P}_{45}$, which has important practical implications, is whether it is \emph{starshaped}.
A set ${\cal A}$ is called starshaped if it contains a point $x \in {\cal A}$ such that all points in $\mathcal{A}$ are visible from $x$.
The point $x$ is then called a \emph{star center} of $\mathcal{A}$. We refer to ref.\ \cite{Hansen:2020} for a detailed overview of properties of starshaped sets.

We are going to prove that ${\cal P}_{45}$ is not starshaped. We will assume that ${\cal P}_{45}$ is starshaped and show that this assumption leads to a contradiction.
The domain ${\cal P}_{45}$ owes its non-trivial geometry to the quartic constraint $\Delta_5 <0$. The remaining constraints in~\cref{eq:P45} are in fact linear, so if they are satisfied at points $X_1$ and $X_2$ then they are also satisfied along the line segment connecting $X_1$ and $X_2$. For this reason we will focus on the quartic constraint $\Delta_5 <0$ hereafter.

Let us assume that ${\cal P}_{45}$ is starshaped and let the point $X_0 \in {\cal P}_{45}$ be a star center. Since the inequalities defining ${\cal P}_{45}$ are homogeneous, we parametrize $X_0$ without loss of generality as
\begin{align}
X_0 := \left( p_1^{2\, (0)} = 1 , s_{12}^{(0)} ,\, s_{23}^{(0)} ,\, s_{34}^{(0)} , \, s_{45}^{(0)} ,\, s_{15}^{(0)} \right) \,. \label{eq:X0gen}
\end{align}
We constrain $X_0$ by requiring that certain points of ${\cal P}_{45}$ are visible from it. 

Let us first consider the one-parameter family of phase-space points
\begin{align}
X^{(z)}_{\rm Regge} := \left( p_1^2 = 1,\, s_{12} = 2 ,\, s_{23} = 3 ,\, s_{34} = -1 , \, s_{45} = z ,\, s_{15} = -1 \right) \,. \label{eq:Xz}
\end{align}
One can easily check that the inequalities~\p{eq:P45} are satisfied for $z \gg 1$. Physically, the asymptotics $z \gg 1$ is the Regge limit with high energies $s_{13} \sim s_{45} \sim z$. We connect $X_0$ and $X^{(z)}_{\rm Regge}$ by a line segment $\gamma_z(t)$ parameterized with $0\leq t \leq 1$. Along the line segment the leading term of $\Delta_5$ for $z \gg 1$ is 
\begin{align} \label{eq:DeltaRegge}
\Delta_5(\gamma_{z}(t)) = t^2(1-t)^2 \left( s_{15}^{(0)}-s_{34}^{(0)} \right)^2 z^2 + {\cal O}(z) \,.
\end{align}
The requirement that $X^{(z)}_{\rm Regge}$ is visible from $X_0$ implies the negativity of $\Delta_5(\gamma_{z}(t))$ for all $t\in[0,1]$. Consequently, the leading term in the $z$-expansion~\eqref{eq:DeltaRegge} has to vanish, giving the following constraint on $X_0$,
\begin{align} \label{eq:s15s34baseconstraint}
s_{15}^{(0)} = s_{34}^{(0)} \,.
\end{align}

In order to obtain more constraints on $X_0$ we consider those permutations of the one-parameter family~\p{eq:Xz} which are automorphisms of the domain ${\cal P}_{45}$, namely the permutations in the set $\Sigma\left(\mathcal{P}_{45}\right)$~\eqref{eq:automorphisms}.
Since $\Delta_5$ is invariant under all $S_4$ permutations, we automatically obtain three more constraints on $X_0$:
\begin{align}\label{eq:sijbaseconstraint}
s_{14}^{(0)} = s_{35}^{(0)} \,,\qquad
s_{15}^{(0)} = s_{24}^{(0)} \,,\qquad
s_{14}^{(0)} = s_{25}^{(0)} \,,
\end{align}
where we tacitly imply that $s_{ij}^{(0)}$ denotes a linear function of the five independent parameters from \p{eq:X0gen}.
Interestingly, the constraints in~\cref{eq:s15s34baseconstraint,eq:sijbaseconstraint} imply that, if $\mathcal{P}_{45}$ is starshaped,
its star centers are in the locus where all linear spurious letters $\mathcal{S}_{\ell}$~\eqref{eq:spurious_linear} vanish.

Combining together the four constraints on $X_0$ we restrict it to the following one-parameter family,
\begin{align}
X_0 = ( p_1^2 = 1, \, s_{12} = a+2, \, s_{23} = a+1, \, s_{34} = -a-1, \, s_{45} = 3a+4, \, s_{15} = -a-1 ) \,, \label{eq:X0a}
\end{align}
with $a>0$.
Note that the base point $X_0$~\p{eq:basepoint} we use in our implementation of the one-mass pentagon functions is exactly of this form with $a = 1$.

Finally we present another one-parameter family of phase-space points $X^{(z)}_{\rm col} \in {\cal P}_{45}$ which is not visible from $X_0$ for any $a>0$ at $z \ll 1$, 
\begin{align}
X^{(z)}_{\rm col} =  \left( p_1^2 = 8 , s_{12} = 10 ,\, s_{23} = z ,\, s_{34} = -1 , \, s_{45} = 12 ,\, s_{15} = -2 \right)\,.
\end{align}
This asymptotics corresponds physically to the collinear limit $p_2 || p_3$.
We connect $X_0$ and $X_{\rm col}^{(z)}$ by a line segment $\gamma_z(t) = (1-t)\, X_0 + t\, X_{\rm col}^{(z)}$. One can then verify that the quadratic constraint $\Delta_5 <0$ cannot be satisfied for all $t\in [0,1]$ for $z\ll 1$. We do this by showing that, for $z \ll 1$ and for any $a > 0$, the sign of $\Delta_5(\gamma_z(t))$ changes at some value of $t \in (0,1)$. For this purpose we find it convenient to pull back $\Delta_5$ to $[0,+\infty)$ by the map $t = \frac{u}{1+u}$ with $u \geq 0$,
\begin{align}
\Delta_5\left(\gamma_z\left( \frac{u}{1+u}\right)\right) = - \frac{(1+a)}{(1+u)^4}(4+3a +12 u) \left((1+a-2u)^2 - (a+1)\right) + {\cal O}(z) \,.
\end{align}
Clearly, the sign of the leading term of this expansion changes as $u$ varies along the positive axis for any value of $a$. In other words, $\Delta_5$ cannot be negative along the entire line for any choice of $a > 0$, and $X_{\rm col}^{(z)}$ is not visible from any $X_0$ of the form~\p{eq:X0a}. Therefore we can conclude that ${\cal P}_{45}$ is not starshaped.

\section{Permutation closure of the planar alphabet}
\label{app:alphabet}

In this appendix we spell out the expressions of the relevant letters $\mathbb{A}_{S_4}^{{\rm rel}}$ listed in~\cref{eq:fullalphabet}. Following ref.~\cite{Abreu:2021smk}, we present them grouped into orbits of the permutation group $S_4$ starting from a generating set of letters. The generating letters may be invariant under certain permutations, so that only a subset of $S_4$ is required to generate the full orbit. We give in the ancillary file \texttt{alphabet\_permutation\_orbits.m}~\cite{supplemental} the subsets of permutations which, starting from the generating letters shown here, give the letters in the corresponding orbits in the correct ordering.

We also highlight the behavior of each letter with respect to changing the sign of one of the square roots. We recall that a letter $W_i$ is called odd with respect to a square root $\sqrt{\delta}$ if $\dd \log W_i$ changes sign when changing the sign of $\sqrt{\delta}$, namely if $W_i$ goes to its inverse $1/W_{i}$. The letters with non-trivial behavior with respect to changing the sign of the square roots are listed in table~\ref{tab:odd_letters}. 

The purely rational relevant letters are given by\footnote{The letters $\{W_{19},\ldots,W_{24}\}$ require also a factor of $-1$ on top of the permutation of the generating letter.}
\begin{align}
\begin{aligned}
W_1 & = p_1^2\,, \\ 
\{W_{2},\ldots, W_{5}\} & =  S_4 \circ \left(s_{12}\right) \,, \\
\{W_{6},\ldots, W_{11}\} & =  S_4 \circ \left(s_{23}\right) \,, \\
\{W_{12},\ldots, W_{15}\} & =  S_4 \circ \left(s_{12}-p_1^2\right)    \,, \\
\{W_{16},\ldots, W_{27}\} & = S_4 \circ \left(s_{15}-s_{34} \right)    \,, \\
\{W_{28},\ldots, W_{33}\} & = S_4 \circ \left(s_{12} s_{15}-p_1^2 s_{34} \right)    \,, \\
\{W_{34},\ldots, W_{45}\} & = S_4 \circ \left( s_{12} s_{23}+p_1^2 s_{45}-s_{12} s_{45} \right)    \,, \\
\{W_{46},\ldots, W_{57}\} & = S_4 \circ \left( s_{15} s_{45} + p_1^2 s_{23} - p_1^2 s_{15} \right)    \,, \\
\{W_{70},\ldots, W_{93}\} & = S_4 \circ \left(s_{12} s_{15}-s_{12} s_{23}-p_1^2 s_{34} \right)    \,,
\end{aligned}
\end{align}
where $S_4 \circ x$ denotes the $S_4$-orbit of $x$, namely the set $\{\sigma \circ x \, | \, \sigma \in S_4\}$.

The letters with non-trivial behaviour with respect to changing the sign of one of the three-mass-triangle square roots but independent of $\trfive$ are
\begin{align}
\begin{aligned}
\{W_{118},\ldots, W_{123}\} & = S_4 \circ \left(\frac{p_1^2-s_{23}+s_{45}+\sqrt{\Delta_3^{(1)}}}{p_1^2-s_{23}+s_{45}-\sqrt{\Delta_3^{(1)}}} \right)   \,, \\
\{W_{124},\ldots, W_{129}\} & =  S_4 \circ \left( \frac{s_{13}-s_{12}-\sqrt{\Delta_3^{(1)}}}{s_{13}-s_{12}+\sqrt{\Delta_3^{(1)}}} \right)   \,. \\
\end{aligned}
\end{align}
In particular, the letters $\{W_{118}, W_{123}, W_{124}, W_{129}\}$, $\{W_{119}, W_{122}, W_{125}, W_{128}\}$ and \linebreak
$\{W_{120}, W_{121}, W_{126}, W_{127}\}$ are odd with respect to $\sqrt{\Delta_3^{(1)}}$, $\sqrt{\Delta_3^{(2)}}$ and $\sqrt{\Delta_3^{(3)}}$, respectively.

The parity-odd letters ---~namely those whose $\dd \log$ changes sign when changing the sign of $\trfive$\footnote{\label{note1}In ref.~\cite{Abreu:2021smk} the letters are defined in terms of $\sqrt{\Delta_5}$, rather than $\trfive$, and are thus even under parity. We adopt their definition of the letters but replace $\sqrt{\Delta_5}$ with $\trfive$, to match the conventions used in the computation of the planar integral families~\cite{Abreu:2020jxa}. Since $\sqrt{\Delta_5}$ is invariant under any permutation of the massless legs whereas $\trfive$ changes sign with an odd-signature permutation, the sets of permutations used here may differ from those of ref.~\cite{Abreu:2021smk}.}~--- which are free from the three-mass-triangle square roots are 
\begin{align} \label{eq:W_130_137}
\{W_{130},\ldots, W_{137}\} = S_4 \circ \left( \frac{a-\trfive}{a+\trfive} \right)   \,,
\end{align}
with
\begin{align}
a = s_{12} s_{23}+s_{23} s_{34}-s_{34} s_{45}+s_{45} s_{15}-s_{12} s_{15}+p_1^2 s_{34} \,.
\end{align}

There are three parity-odd letters which also depend on one of the three-mass-triangle square roots,
\begin{align} \label{eq:W_186_188}
\begin{aligned}
W_{186} & = \frac{\Omega^{--}\Omega^{++}}{\Omega^{-+}\Omega^{+-}} \,, \\
W_{187} & = \sigma \circ W_{186}  \quad \text{with} \ \sigma = \left(2435\right) \,, \\
W_{188} & = \sigma \circ \left(\frac{1}{W_{186}}\right)  \quad \text{with} \ \sigma = \left(2543\right) \,, \\
\end{aligned}
\end{align}
where
\begin{align}
\Omega^{\pm \pm} = s_{12} s_{15} - s_{12} s_{23} - s_{15} s_{45} \pm s_{34} \sqrt{ \Delta_3^{(1)} } \pm \trfive \,.
\end{align}
The letters $W_{186}$, $W_{187}$ and $W_{188}$ are all odd with respect to $\trfive$, and are odd with respect to $\sqrt{\Delta_3^{(1)}}$, $\sqrt{\Delta_3^{(2)}}$ and $\sqrt{\Delta_3^{(3)}}$, respectively. The definition of $W_{188}$ in~\cref{eq:W_186_188} requires an inverse because we are using $\trfive$ rather than $\sqrt{\Delta_5} \,$ (see also \cref{note1}).

The last four relevant letters are given by the four square roots of the problem,
\begin{align}
\begin{aligned}
W_{195} & = \sqrt{\Delta_3^{(1)}} \,, \\ 
W_{196} & = \sqrt{\Delta_3^{(2)}} \,, \\
W_{197} & = \sqrt{\Delta_3^{(3)}} \,, \\
W_{198} & = \trfive \,.
\end{aligned}
\end{align}
We stress that these letters are even with respect to changing the sign of any square root.

\begin{table}
\centering
\begin{tabular}{cl}
\toprule 
Square root & Odd letters \\
\midrule
$\sqrt{\Delta_3^{(1)}}$ & $W_{118}\,, W_{123}\,, W_{124}\,, W_{129}\,, W_{186}$ \\
$\sqrt{\Delta_3^{(2)}}$ & $W_{119}\,, W_{122}\,, W_{125}\,, W_{128}\,, W_{187}$ \\
$\sqrt{\Delta_3^{(3)}}$ & $W_{120}\,, W_{121}\,, W_{126}\,, W_{127}\,, W_{188}$ \\
\vphantom{$\sqrt{\Delta_3^{(3)}}$} $\trfive$ & $W_{130}\,, \ldots \,, W_{137}\,, W_{186}\,, W_{187}\,, W_{188}$ \\
\bottomrule
\end{tabular}
\caption{Relevant letters which depend on the sign of the square roots of the problem. On each row the letters in the right column are odd with respect to changing the sign of the square root in the left column. We recall that $\trfive$ is related to $\sqrt{\Delta_5}$ through~\cref{eq:tr5Delta5}.}
\label{tab:odd_letters}
\end{table}

\section{Positivity of the alphabet in the physical scattering region $45 \to 123$}
\label{app:alphpositivity}

The one-mass pentagon functions are expressed as iterated integrals with integration kernels given by the logarithms of the alphabet letters (see section~\ref{sec:IteratedIntegrals}). The representation in terms of iterated integrals highlights that the pentagon functions may have singularities only on the loci where any of the involved letters vanishes or diverges.
Therefore we need to inquire about the \textit{positivity} of the alphabet letters in the region ${\cal P}_{45}$, i.e.\ whether they have definite sign in the region ${\cal P}_{45}$.
This has important implications for the construction of the explicit representation of the pentagon functions in \cref{sec:explicit-repr}.
In this appendix we study systematically the positivity of the $108$ relevant letters of the alphabet listed in~\cref{eq:fullalphabet}. 
We show that $20$ of them do not have fixed sign in ${\cal P}_{45}$. We call the latter {\em spurious letters}. Out of them, $4$ are linear in the Mandelstam invariants,
\begin{align} \label{eq:app_spurious_linear}
{\cal S}_{\ell} = \{ & W_{16},\,W_{17},\, W_{19},\,W_{20} \} \,,
\end{align} 
and $16$ are quadratic,
\begin{equation} \label{eq:app_spurious_quadratic}
\begin{aligned}
{\cal S}_q = \{ & W_{35},\,W_{36},\,W_{38},\,W_{39},\,W_{40},\,W_{41},\,W_{43},\,W_{44},\,\\
& W_{72},\,W_{74},\,W_{78},\,W_{80},\,W_{82},\,W_{84},\,W_{88},\,W_{90} \} \,.
\end{aligned}
\end{equation} 
We denote by ${\cal S}$ the set of all spurious letters,
\begin{align}
{\cal S} = {\cal S}_{\ell}  \cup {\cal S}_q \subset \mathbb{A}_{S_4}^{{\rm rel}}  \,.
\end{align}
We prove this by rewriting the letters in such a way that their positivity properties in ${\cal P}_{45}$ become obvious consequences of the inequalities~\p{eq:P45}. 
For the sake of presentation we split the alphabet into several subsets of letters having similar structure, to each of which we devote a subsection. Finally, in subsection~\ref{sec:3massTriangles} we show that the three-mass triangle functions defined in~\cref{eq:tri-w2} have no branch cuts within ${\cal P}_{45}$.

\subsection{Linear letters}

The first $27$ letters, $\{ W_{i} \}_{i =1}^{27}$, are linear in the Mandelstam variables~\p{eq:Mand}. By rewriting them as
\begin{equation}
\begin{alignedat}{3}
& W_1= p_1^2 \,, && W_2 = s_{12} \,, && W_3= s_{13}\,, \\
& W_4 = s_{14}\,, && W_{5} = s_{15} \,, && W_6 = s_{23}\,, \\
& W_7 = s_{24} \,, && W_8 = s_{25}\,, && W_{9} = s_{34}\,, \\
& W_{10} = s_{35} \,, && W_{11} = s_{45}\,, && W_{12} = s_{12}-p_1^2 \,, \\
& W_{13} = s_{13}-p_1^2 \,, && W_{14} = s_{14}-p_1^2 \,, && W_{15} = s_{15}-p_1^2 \,, \\ 
& W_{16} = s_{23}+s_{24}\,, && W_{17} = s_{23}+s_{25} \,, && W_{18} = s_{24}+s_{25} \,, \\
& W_{19} = -s_{23}-s_{34} \,, && W_{20} = -s_{23}-s_{25} \,, && W_{21} = (s_{13}-p_1^2)+s_{23} \,, \\
& W_{22} = -s_{24}-s_{34} \,, \quad \quad && W_{23} = (s_{14}-p_1^2)+s_{34} \,,  \quad \quad && W_{24} = (s_{14}-p_1^2)+s_{24}\,, \\
& W_{25} = s_{25}+s_{35} \,, && W_{26} = (p_1^2-s_{15})-s_{35} \,, && W_{27} = (p_1^2-s_{15})-s_{25} \,,
\end{alignedat}
\end{equation}
we can immediately deduce their positivity properties in ${\cal P}_{45}$ from the defining inequalities~\eqref{eq:P45}. Only the inequalities which are linear in the Mandelstam invariants are required. The positivity properties of the linear letters are summarized as follows:
\begin{equation}
\begin{tabular}{rm{8ex}}
 $W_{1},\,W_{2},\,W_{3},\,W_{6},\,W_{11},\,W_{12},\,W_{13},\,W_{21},\,W_{22},\,W_{23},\,W_{26},\,W_{27}$ & $>0$ \\[1em]
 $W_{4},\,W_{5},\,W_{7},\,W_{8},\,W_{9},\,W_{10},\,W_{14},\,W_{15},\,W_{18},\,W_{24},\,W_{25}$ & $<0$ \\[1em]
 $W_{16},\,W_{17},\, W_{19},\,W_{20}$ & $\gtrless 0$ \\ 
\end{tabular}
\end{equation}

\subsection{Quadratic letters}

$54$ letters, $\{ W_{i} \}_{i = 28}^{57} \cup \{ W_{i} \}_{i = 70}^{93}$, are quadratic in the Mandelstam variables \p{eq:Mand}.
For $11$ of them we find a representation which makes their positivity a manifest consequence of the linear inequalities~\p{eq:P45} among Mandelstam invariants,
\begin{align} \label{quadPos1}
\begin{aligned}
& W_{49} = s_{35}(s_{13}+p_1^2)+p_1^2(s_{15}-p_1^2)\,,\\
& W_{51} = s_{34}(s_{13}+p_1^2)+p_1^2(s_{14}-p_1^2)\,,\\
& W_{53}=s_{25}(s_{12}+p_1^2)+p_1^2(s_{15}-p_1^2)\,,\\
& W_{55}=s_{24}(s_{12}+p_1^2)+p_1^2(s_{14}-p_1^2)\,,\\
& W_{56} = s_{23}(s_{13}+p_1^2)+p_1^2(s_{12}-p_1^2)\,,\\
& W_{57} =  s_{23}(s_{12}+p_1^2)+p_1^2 (s_{13}-p_1^2)\,,\\
& W_{70} = s_{12} s_{24} + s_{34} (s_{12}-p_1^2) \,,\\
& W_{71} = s_{12} s_{25} + s_{35}(s_{12}-p_1^2)\,,\\
& W_{76} = s_{13} s_{34} +  s_{24}(s_{13}-p_1^2) \,,\\
& W_{77} = s_{13} s_{35} + s_{25} (s_{13} - p_1^2) \,,\\
& W_{93} = s_{15} s_{35} + s_{34} (s_{15}- p_1^2) \,,
\end{aligned}
\end{align}
i.e.\ both terms in each of the previous sums are simultaneously either positive or negative.

In order to establish the positivity of the remaining quadratic letters, the linear inequalities in~\cref{eq:P45} are not sufficient. We have to appeal to the Gram determinant inequalities. We thus express the letters in terms of order-$3$ Gram determinants, $G_{1ij}$, and profit from their positivity in ${\cal P}_{45}$, namely $G_{1ij} >0$ for $i,j=2,\ldots,5$ and $i \neq j$. In this way we can straightforwardly establish that the following quadratic letters do not change sign in ${\cal P}_{45}$,
\begin{align} \label{quadPos2}
\begin{alignedat}{3}
& W_{28} = \frac{4G_{125}}{s_{2 5}} \,, \qquad\qquad\qquad  && W_{29} = \frac{4G_{124}}{s_{24}}\,, \qquad\qquad\qquad && W_{30} = \frac{4G_{123}}{s_{23}} \,, \\
& W_{31} = \frac{4G_{135}}{s_{35}}\,,  && W_{32} = \frac{4G_{134}}{s_{34}} \,,  && W_{33} = \frac{4G_{145}}{s_{45}} \,, \\
& W_{34} = \frac{4G_{124}}{s_{24}} + \frac{4G_{125}}{s_{25}} \,, \qquad\quad && W_{37} = \frac{4G_{134}}{s_{34}}+\frac{4G_{135}}{s_{35}} \,, && \\
& W_{42} = \frac{4G_{124}}{s_{24}}+\frac{4G_{134}}{s_{34}} \,,  && W_{45} = \frac{4G_{125}}{s_{25}} + \frac{4G_{135}}{s_{35}} \,. && \\
\end{alignedat}
\end{align}
We recall that $G_{1ij}/s_{ij}$ is a quadratic polynomial in the Mandelstam invariants, as can be seen explicitly in~\cref{eq:Gord3}.
Similarly, we observe that both terms in the following sums are of the same sign,
\begin{align} \label{quadPos3}
\begin{aligned}
& W_{46} = -\frac{4G_{145}}{s_{45}} + s_{15}(s_{23}-s_{15}) \,,
& W_{47} = -\frac{4G_{145}}{s_{45}} + s_{14}(s_{23}-s_{14}) \,, \\
& W_{73} = -\frac{4G_{125}}{s_{25}} + s_{35}(p_1^2-s_{12})  \,,
& W_{75} = -\frac{4G_{124}}{s_{24}} + s_{34}(p_1^2-s_{12}) \,, \\
& W_{79} = -\frac{4G_{135}}{s_{35}} + s_{25}(p_1^2-s_{13})  \,,
& W_{81} = -\frac{4G_{134}}{s_{34}} + s_{24}(p_1^2-s_{13})  \,, \\  
& W_{83} = -\frac{4G_{145}}{s_{45}} + s_{35}(p_1^2-s_{14}) \,,
& W_{85} = -\frac{4G_{145}}{s_{45}} + s_{25}(p_1^2-s_{14})  \,, \\
& W_{86} = -\frac{4G_{124}}{s_{24}} + s_{23}(p_1^2-s_{14})  \,,
& W_{87} = -\frac{4G_{134}}{s_{34}} + s_{23}(p_1^2-s_{14})  \,,\\
& W_{89} = -\frac{4G_{145}}{s_{45}} + s_{34}(p_1^2-s_{15})  \,, 
& W_{91} = -\frac{4G_{145}}{s_{45}} + s_{24}(p_1^2-s_{15})  \,,\\
& W_{92} = -\frac{4G_{125}}{s_{25}} + s_{23}(p_1^2-s_{15})  \,.
\end{aligned}
\end{align}
Finally, we have to employ more intricate identities to highlight the positivity of the following quadratic letters,
\begin{align} \label{quadPos4}
\begin{aligned}
& W_{48} = \frac{p_1^2 (s_{23}+s_{34})^2 + 4  G_{135} }{s_{13}-p_1^2} \,, \qquad
& W_{50} = \frac{p_1^2 (s_{23}+s_{35})^2 + 4  G_{134} }{s_{13}-p_1^2}  \,, \\
& W_{52} = \frac{p_1^2(s_{15}-s_{34})^2 + 4 G_{125}}{s_{12}-p_1^2} \,, \qquad & W_{54} = \frac{p_1^2 (s_{23}+s_{25})^2 + 4  G_{124} }{s_{12}-p_1^2}  \,. 
\end{aligned}
\end{align}
Let us recall that the letters in the previous equation are polynomial in the Mandelstam variables: the denominators factor out once the numerators are expressed in terms of Mandelstam variables.

We summarize the positivity properties of the quadratic letters given by \cref{quadPos1,quadPos2,quadPos3,quadPos4} as follows:
\begin{equation}
\begin{tabular}{rm{8ex}}
 $\begin{array}{l}W_{30},\,W_{33},\,W_{48},\,W_{50},\,W_{52},\,W_{54},\,W_{56},\,W_{57},\,\\ W_{73},\,W_{75},\,W_{79},\,W_{81},\,W_{86},\,W_{87},\,W_{92},\,W_{93}\end{array}$ &  $>0$ \\[1.5em]
 $\begin{array}{l} W_{28},\,W_{29},\,W_{31},\,W_{32},\,W_{34},\,W_{37},\,W_{42},\,W_{45},\,W_{46},\,W_{47},\,W_{49},\,\\ W_{51},\,W_{53},\, W_{55},\,W_{70},\, W_{71},\,W_{76},\,W_{77},\,W_{83},\, W_{85},\, W_{89},\, W_{91}\end{array}$ & $<0$ \\[1.5em]
 $\begin{array}{l} W_{35},\,W_{36},\,W_{38},\,W_{39},\,W_{40},\,W_{41},\,W_{43},\,W_{44},\,\\
W_{72},\,W_{74},\,W_{78},\,W_{80},\,W_{82},\,W_{84},\,W_{88},\,W_{90}\end{array}$ & $\gtrless 0$ \\ 
\end{tabular}
\end{equation}

\subsection{Three-mass square-root letters}
\label{app:3m-letters}    
    
The letters $W_{195},\, W_{196},\, W_{197}$ are the three-mass triangle square roots, $\sqrt{\Delta^{(i)}_3}$ with $i=1,2,3$. They are given by order-$2$ Gram determinants~\p{Deltalambda}, and thus they are positive in ${\cal P}_{45}$ due to the proposition in~\cref{eq:idG3},  
\begin{align}
W_{195} ,\, W_{196} ,\, W_{197} > 0 \,.
\end{align}

Among the square-root letters, the letters $\{W_{i} \}_{i=118}^{129}$ contain only the square roots of three-mass-triangle type. We observe that they are neither singular nor vanish in the physical region ${\cal P}_{45}$, and
their signs are summarized as follows:
\begin{equation}
\begin{tabular}{rm{8ex}}
  $W_{118},\,W_{123},\,W_{125},\,W_{126},\,W_{127},\,W_{128}$ & $>0$ \\[1em]
  $W_{119},\,W_{120},\,W_{121},\,W_{122},\,W_{124},\,W_{129}$ & $<0$ \\ 
\end{tabular}
\end{equation}
In order to show this, let us consider two prototypical examples. 

The letter $W_{118}$,
\begin{align}
W_{118} = \frac{s_{12} + s_{13} + \sqrt{\Delta^{(1)}_3}}{s_{12} + s_{13} - \sqrt{\Delta^{(1)}_3}}\,,
\end{align}
could become singular or vanish only at $(s_{12}+s_{13})^2 = \Delta_3^{(1)}$. The latter constraint is equivalent to $p_1^2 \, s_{45} = 0$, which is impossible inside the region ${\cal P}_{45}$. Therefore, $W_{118}$ cannot change sign in ${\cal P}_{45}$.

The second example is given by the letter $W_{126}$,
\begin{align}
W_{126} = \frac{s_{12} - s_{15} + \sqrt{\Delta^{(3)}_3}}{s_{12} - s_{15} - \sqrt{\Delta^{(3)}_3}}\,.
\end{align}
The latter vanishes or becomes singular only at $(s_{12}-s_{15})^2 = \Delta_3^{(3)}$, which is equivalent to $s_{12}s_{15} = p_1^2 s_{34}$. Resolving the previous relation in one of the Mandelstam variables, e.g.\ $p_1^2 = s_{12}s_{15}/s_{34}$, and eliminating this variable from $\Delta_5$, we observe that $\Delta_5$ turns into a perfect square,
\begin{align}
\Delta_5\Bigl|_{p_1^2 = \frac{s_{12}s_{15}}{s_{34}}} = (s_{12}s_{23}-s_{23}s_{34}-s_{15}s_{45}+s_{34}s_{45})^2 \,,
\end{align}
which cannot be negative, in contradiction with the quadratic constraint $\Delta_5 <0$ in~\cref{eq:P45}.

For the remaining algebraic letters involving only the three-mass triangle square roots, one of the two arguments above is applicable.

\subsection{Parity-odd letters and $\sqrt{\Delta_5}$} 
\label{app:c-val-letters}

The remaining relevant letters that we need to consider, $\{W_{i}\}_{i=130}^{137} \cup \{W_{i}\}_{i=186}^{188} \cup \{ W_{198} \}$, all contain $\sqrt{\Delta_5}$ in their expression. Since $\Delta_5 < 0$ in ${\cal P}_{45}$, $\sqrt{\Delta_5}$ is non-vanishing and purely imaginary in the physical region. Thus, $W_{198} =\sqrt{\Delta_5}$ is purely imaginary, while the other eleven letters take complex values on the unit circle. The latter are parity-odd because the sign of the corresponding $\dd \log$'s changes under the action of space-time parity $P$,
\begin{align}
P : \ \dd \log W_i \to - \dd  \log W_i \quad \text{for} \quad  i = 130,\ldots,137,186,187,188 \,.
\end{align}
Moreover, these letters cannot be equal to $1$, as that would imply $\Delta_5 = 0$. In summary, for the parity-odd letters we have that
\begin{align}
|W_i| = 1 \quad \text{and} \quad  W_i \neq 1 \quad  \text{for} \quad i = 130,\ldots,137,186,187,188\,. 
\end{align}

\subsection{Three-mass triangles}
\label{sec:3massTriangles}

In~\cref{eq:tri-w2} we defined three weight-two pentagon functions corresponding, up to an overall algebraic factor, to the finite part of the three-mass triangle Feynman integrals. Their explicit expressions~\eqref{eq:tri-w2} rely on the representation~\p{eq:trifun} of the three-mass triangle function. Here we show that they are well defined, namely real analytic, in the physical channel ${\cal P}_{45}$.

The following inequalities,
\begin{align}
s_{45} - s_{23} + p_1^2 \pm \sqrt{\Delta_3^{(1)}} > 0  \,,\qquad 
s_{45}+s_{23} - p_1^2 \pm \sqrt{\Delta_3^{(1)}} > 0 \,, \label{eq:C11}
\end{align}
imply that the arguments of the logarithms and dilogarithms in the expression for $f^{(2)}_{23}$ do not cross any branch cuts. The strict inequalities~\p{eq:C11} turn into equalities at the boundaries $p_1^2 s_{45} = 0$ and $s_{23} s_{45} = 0$ of ${\cal P}_{45}$, respectively.
Similarly, the inequalities
\begin{align}
\begin{aligned}
& \pm(p_1^2 + s_{34}-s_{25})+\sqrt{\Delta_3^{(3)}} > 0 \,, \\
& \pm (p_{1}^2 + s_{25}-s_{34}) + \sqrt{\Delta_3^{(3)}} > 0 \,, \\
& p_{1}^2-s_{25}-s_{34} \pm \sqrt{\Delta_3^{(3)}} > 0  \,,
\end{aligned}
\end{align}
guarantee absence of branch cuts in ${\cal P}_{45}$ for $f^{(2)}_{24}$. The same holds true for $f^{(2)}_{25}$, thanks to
\begin{align}
\begin{aligned}
& \pm (p_1^2 + s_{24}-s_{35}) +\sqrt{\Delta_3^{(2)}} > 0 \,, \\
& \pm (p_1^2 - s_{24} + s_{35}) +\sqrt{\Delta_3^{(2)}} > 0 \,, \\
& p_1^2 - s_{24}-s_{35} \pm\sqrt{\Delta_3^{(2)}} > 0 \,.
\end{aligned}
\end{align}

\bibliography{pentagon_functions_1m.bib}
\bibliographystyle{JHEP}

\end{document}